\documentclass[aps,prb,showpacs,twocolumn,floats,floatfix]{revtex4}

\usepackage{epsfig,amssymb}

\newcommand\beq{\begin{equation}}
\newcommand\eeq{\end{equation}}
\newcommand\bea{\begin{eqnarray}}
\newcommand\eea{\end{eqnarray}}
\newcommand\non{\nonumber}
\newcommand\noi{\noindent}
\newcommand\al{\alpha}

\newcommand\de{\delta}
\newcommand\ep{\epsilon}

\newcommand\omc{\omega_c}
\newcommand\Omc{\Omega_c}
\newcommand\dg{\dagger}

\newcommand\ig{\includegraphics}
\newcommand\bib{\bibitem}

\begin{document}

\textheight=23.8cm

\title{Quasi-particle propagation in quantum Hall systems}
\author{Diptiman Sen$^1$}
\author{Michael Stone$^2$}
\author{Smitha Vishveshwara$^2$}
\affiliation{$^1$ Centre for High Energy Physics, Indian Institute of Science,
Bangalore 560012, India \\
$^2$ Department of Physics, University of Illinois at Urbana-Champaign, 1110
W. Green St., Urbana Illinois 61801-3080, USA}

\date{\today}

\begin{abstract}
We study various geometrical aspects of the propagation of particles obeying
fractional statistics in the physical setting of the quantum Hall system. We
find a discrete set of zeros for the two-particle kernel in the lowest Landau
level; these arise from a combination of a two-particle Aharonov-Bohm effect
and the exchange phase related to fractional statistics. The kernel also shows
short distance exclusion statistics, for instance, in a power law behavior as a
function of initial and final positions of the particles. We employ the
one-particle kernel to compute impurity-mediated tunneling amplitudes between
different edges of a finite-sized quantum Hall system and and find that they
vanishes for certain strengths and locations of the impurity scattering
potentials. We show that even in the absence of scattering, the correlation
functions between different edges exhibits unusual features for a narrow enough
Hall bar.
\end{abstract}
\maketitle
\vskip .6 true cm

\section{Introduction}

Fractional statistics in two dimensions \cite{leinaas,wilczek1} has been
studied extensively ever since it was proposed that quasi-particles in
fractional quantum Hall systems carry this attribute.
\cite{laughlin1,laughlin2,halperin,arovas,jain1} Such particles are called
anyons; the wave function of two or more such particles picks up a phase (which
may be different from $\pm 1$) when any two of them are exchanged. Interest in
this subject has grown further in recent years due to the possibility of
topological quantum computation \cite{kitaev1} and the development of spin
models which support anyonic excitations \cite{kitaev2,pachos,han,chen}.
Several proposals
\cite{kivelson,jain2,chamon1,isakov,safi,kane1,vishveshwara1,kim,grosfeld,law1,braggio,feldman}
and experimental attempts \cite{camino} have been made to detect the fractional
statistics of quasi-particles in fractional quantum Hall systems. (Here we
restrict our attention to Abelian fractional statistics). While there are many
studies of anyons propagating at the edges of fractional quantum Hall systems
\cite{wen1}, anyon correlations in the bulk have not been studied in as much
detail \cite{martin}. Since anyons are intrinsically two-dimensional objects, 
a coherent understanding of their behavior necessitates a study of their
properties in the bulk. Here we undertake such a study of the bulk properties
of anyons in the lowest Landau level of a quantum Hall system.

Given that fractional statistics is an intrinsically two-particle
concept, the most direct way of understanding anyonic behavior is by
investigating the properties of the two-particle kernel. Historically,
two-particle kernels have played an important role in many areas of physics
including scattering problems in particle physics, quantum optics and
astrophysics \cite{baym,hbt}. In particular, seminal work by Hanbury Brown and
Twiss showed that bosons tend to `bunch' when arriving at a given point in
space-time. Fermions, on the other hand, tend to `anti-bunch'. Two-particle
kernels are in fact key for studying the effects of quantum statistics on
correlations between two identical particles which are propagating together.
The one-particle kernel is also useful to study even though it is not directly
affected by the exchange phase; for instance, it provides information about the
charge of the particle (if one considers its motion in a magnetic field) and
tunneling between different edges of a quantum Hall system. Moreover, as
elucidated in what follows, information on exchange statistics can also be
derived by analyzing the extent to which two-particle properties can be
decomposed into single-particle properties.

In this paper, we analyze several aspects of the propagation of anyons through
the bulk of a quantum Hall system in the presence of a strong magnetic field.
In particular, we show that the simultaneous propagation of two anyons in the
lowest Landau level exhibits some surprising features depending on the
geometrical arrangement of the initial and final positions. These arise from an
interplay of the Aharonov-Bohm phase due to the magnetic field and the phase
due to the exchange of two anyons. We discuss a physical realization of these
anyon trajectories. As a first step towards connecting bulk physics to the
boundary of the quantum Hall system, as is relevant to any physical situation,
we then analyze single-particle properties in a bounded system. We discuss some
unusual features which can arise when a single particle propagates from one
edge to another through the bulk of a quantum Hall system which has the form
of a narrow strip.

The plan of this paper is as follows. In Sec. II, we discuss the one-particle
kernel of a charged particle moving in a magnetic field. This will be used
later to compute the two-particle kernel (using centre of mass and relative
coordinates) and the amplitude for tunneling across a narrow quantum Hall
system. In Sec. III, we calculate the two-particle kernel for fermions, bosons
and anyons in general. We show that the two-anyon kernel in the lowest Landau
level vanishes if the initial and final positions satisfy a special set of
conditions. These conditions can be given a simple interpretation in terms of
the phase difference of two paths. We propose some experiments for using the
two-anyon kernel to measure their charge and exchange statistics. The possible
corrections arising from higher Landau levels are also discussed.
In Sec. IV, we consider tunneling of a particle between different edges
of a quantum Hall strip; the tunneling is induced by impurities
which may be present anywhere in the strip. We show that the tunneling
amplitude vanishes if the strengths and locations of the impurities satisfy
certain conditions. We propose a Hall bar geometry for observing these
effects. In Sec. V, we study two-point correlators along different
edges of a quantum Hall system. Once again, we find that the correlator can
vanish if the system has the form of a narrow strip, and the two points are
separated by some particular distances. We summarize our results in Sec. VI.
A condensed version of this work has appeared earlier \cite{vishveshwara2}.

In the discussions to follow, various correlators are analyzed depending on
the properties in question. We clarify their definitions at the outset: since
we will discuss several correlators in this paper, it may be useful to list
them here. In Secs. II and III, we will discuss the one- and two-particle
kernels. The one-particle kernel $K_1 ({\vec r}_f ; {\vec r}_i; t)$ is the
amplitude for a particle to go from an initial position ${\vec r}_i$ to a final
position ${\vec r}_f$ in time $t$, where $t > 0$. We define the kernel to be
zero if $t < 0$. Given a vacuum state $|0 \rangle$, and a second quantized
field $\Psi ({\vec r}, t)$ which annihilates a particle at $({\vec r}, t)$, the
kernel is given by \bea K_1 ({\vec r}_f; {\vec r}_i; t) &=& \langle 0 | \Psi
({\vec r}_f, t)
\Psi^\dg ({\vec r}_i , 0) | 0 \rangle \non \\
&=& \sum_n ~\psi_n ({\vec r}_f) ~\psi_n^* ({\vec r}_i) ~e^{-iE_n t/ \hbar},
\label{eqn1} \eea
where $E_n$ and $\psi_n$ in the second line denote the eigenvalues and
eigenfunctions respectively of the one-particle Hamiltonian, and we have
assumed the label $n$ to be discrete. The Fourier transform of this kernel
is given by
\bea {\tilde K}_1 ({\vec r}_f; {\vec r}_i; E) &=& \int_0^\infty ~dt~ K_1
({\vec r}_f; {\vec r}_i; t) ~e^{iEt/\hbar} \non \\
&=& -i \hbar ~\sum_n ~\frac{\psi_n ({\vec r}_f) \psi_n^* ({\vec r}_i)}{E_n - E
- i \ep}. \eea
The two-particle kernel is defined in an analogous way. In Sec.
IV, we will use the retarded Green's function; this is related to the
one-particle kernel as $G ({\vec r}_f; {\vec r}_i; E) = (i /\hbar) {\tilde K}_1
({\vec r}_f; {\vec r}_i; E)$. Finally, in Sec. V, we will discuss a two-point
correlator. Given the ground state $|G \rangle$ of a system of several
non-interacting electrons in which all states up to a Fermi energy $E_F$ are
filled (the case of interest in Sec. V), we will be interested in the
correlator
\bea C ({\vec r}_1; {\vec r}_2; t) &=& \langle G | \Psi^\dg ({\vec r}_1, 0)
\Psi ({\vec r}_2, t)| G \rangle \non \\
&=& \sum_{E_n < E_F} ~\psi_n^* ({\vec r}_1) ~\psi_n ({\vec r}_2) ~e^{-iE_n
t/ \hbar}. \label{eqn3} \eea
Note that Eqs. (\ref{eqn1}) and (\ref{eqn3}) differ in the range of values
of $n$ which is summed over.

\section{One-particle kernel}

For a free particle moving in two dimensions, say the $x-y$ plane, the
one-particle kernel is given by
\beq K_1 ({\vec r}_f; {\vec r}_i; t) ~=~ \frac{\mu}{i2\pi \hbar t} ~e^{i
\mu ({\vec r}_f - {\vec r}_i)^2 /(2\hbar t)}, \label{k1} \eeq
where $\mu$ is the mass of the particle.

Let us now consider the effect of a uniform magnetic field ${\vec B} = B
{\hat z}$ applied perpendicular to the $x-y$ plane. If we use the symmetric
gauge ${\vec A} = (1/2) {\vec B} \times {\vec r}$, the Hamiltonian is given by
\beq H ~=~ \frac{1}{2\mu} ~(p_x + \frac{eB}{2c} y)^2 ~+~ \frac{1}{2\mu} ~(p_y
- \frac{eB}{2c} x)^2, \label{ham1} \eeq
where $e$ and $c$ denote the charge of the particle and the speed of light
respectively. For convenience, we assume that $eB > 0$. The cyclotron
frequency is given by $\omc = eB/\mu c$. As shown in Appendix A, the kernel
is given by
\bea & & K_1 ({\vec r}_f; {\vec r}_i; t) \non \\
& & = ~\frac{eB}{i 4\pi \hbar c \sin (\omc t/2)} ~\exp [~ \frac{ieB \cot
(\omc t/2)}{4 \hbar c} ~({\vec r}_f - {\vec r}_i)^2 \non \\
& & ~~~~~~~~~~~~~~~~~~~~~~~~~~~~~~~+~ \frac{ieB}{2 \hbar c} ~{\hat z} \cdot
{\vec r}_i \times {\vec r}_f ~]. \label{k1mag1} \eea

Eq. (\ref{k1mag1}) can be written in imaginary time by replacing $t$ by
$-i\tau$. We then obtain
\bea & & K_1 ({\vec r}_f; {\vec r}_i; \tau) \non \\
& & = \frac{eB}{4\pi \hbar c \sinh (\omc \tau/2)} ~\exp [ -~
\frac{eB \coth (\omc \tau /2)}{4 \hbar c} ({\vec r}_f - {\vec r}_i)^2 \non \\
& & ~~~~~~~~~~~~~~~~~~~~~~~~~~~~~~~+~ \frac{ieB}{2 \hbar c} ~{\hat z} \cdot
{\vec r}_i \times {\vec r}_f ]. \label{k1mag2} \eea
The kernel in the lowest Landau level (LLL) can be obtained by multiplying
Eq. (\ref{k1mag2}) by $e^{\omc \tau/2}$ (this is equivalent to shifting all
the energy levels by $\hbar \omc/2$ in order to make the ground state
energy zero), and then letting $\tau \to \infty$,
\bea K_1 ({\vec r}_f; {\vec r}_i) &=& \frac{eB}{2\pi \hbar c} ~\exp
[ -~\frac{eB}{4 \hbar c} ~({\vec r}_f - {\vec r}_i)^2 \non \\
& & ~~~~~~~~~~~~~+~ \frac{ieB}{2 \hbar c} ~{\hat z} \cdot
{\vec r}_i \times {\vec r}_f ]. \label{k1mag3} \eea
Absorbing a factor of the Landau length $l = \sqrt{{\hbar c}/eB}$ in
the coordinates ${\vec r}_i$ and ${\vec r}_f$, we have
\bea K_1 ({\vec r}_f; {\vec r}_i)
&=& \frac{1}{2\pi l^2} ~\exp [ -\frac{1}{4} ~({\vec r}_f - {\vec r}_i)^2 +
\frac{i}{2} {\hat z} \cdot {\vec r}_i \times {\vec r}_f ], \non \\
& & \label{k1mag4} \eea
where the positions ${\vec r}_i$ are now dimensionless quantities. We rewrite
Eq. (\ref{k1mag4}) using complex coordinates, $z = x+iy$ and $z^* = x-iy$. We
then obtain
\bea K_1 ({\vec r}_f; {\vec r}_i)
&=& \frac{1}{2\pi l^2} ~\exp [ -\frac{1}{4} ~(|z_f|^2 + |z_i|^2) ~+~
\frac{1}{2} z_f z_i^* ]. \non \\ \label{k1mag5} \eea

Eq. (\ref{k1mag5}) can be derived in a different way. In general, the kernel
is given by a sum over all the eigenstates of the Hamiltonian $H$,
\beq K_1 ({\vec r}_f; {\vec r}_i; \tau) ~=~ \sum_n ~\psi_n ({\vec r}_f) ~
\psi_n^* ({\vec r}_i) ~e^{-E_n \tau /\hbar}. \label{sum} \eeq
The Hamiltonian in Eq. (\ref{ham1}) written in dimensionless complex
coordinates takes the form
\beq H ~=~ \hbar \omc ~[- ~2 \frac{\partial}{\partial z} \frac{\partial}{
\partial z^*} ~+~ \frac{1}{8} zz^* ~-~ \frac{z}{2} \frac{\partial}{
\partial z} ~+~ \frac{z^*}{2} \frac{\partial}{\partial z^*} ]. \eeq
Assuming that the eigenstates of $H$ are of the form $\psi = f(z,z^*) \exp
(- |z|^2/4)$, we obtain the eigenvalue equation ${\tilde H} f = E f$, where
\beq {\tilde H} ~=~ \hbar \omc ~[- ~2 \frac{\partial}{\partial z}
\frac{\partial}{\partial z^*} ~+~ z^* \frac{\partial}{\partial z^*} ~+~
\frac{1}{2} ~]. \label{ham2} \eeq
The ground states of $\tilde H$ are given by analytic functions of $z$.
The normalized wave functions in the LLL are given by
\beq \psi_n ({\vec r}) ~=~ \frac{1}{l \sqrt{n! ~2\pi}} ~\left(
\frac{z}{\sqrt{2}} \right)^n ~\exp [ - \frac{1}{4} ~|z|^2], \label{wave1} \eeq
where $n=0,1,2,\cdots$. For large values of $n$, the probability corresponding
to the $n^{\rm th}$ wave function is peaked on a circle of radius $l
\sqrt{2n}$. Substituting Eq. (\ref{wave1}) in Eq. (\ref{sum}) and letting
$\tau \to \infty$, we obtain the expression in Eq. (\ref{k1mag5}); recall
that we have shifted the energy levels so that $E_n = 0$.

The kernel in Eq. (\ref{k1mag5}) satisfies the reproducing property
\beq \psi ({\vec r}_f) ~=~ \int ~d^2{\vec r}_i ~K_1 ({\vec r}_f; {\vec r}_i) ~
\psi ({\vec r}_i). \eeq
A simple way to prove this is to note that an arbitrary wave function $\psi$
in the LLL can be written as
\beq \psi ({\vec r}) ~=~ \sum_{n=0}^\infty ~c_n z^n \exp [ -\frac{1}{4}
|z|^2], \eeq
where the $c_n$'s can take any values, and then show that the reproducing
property holds for each term $z^n$ separately.

\section{Two-particle kernel}

\subsection{Fermions and bosons}

For two identical fermions, it is well-known that the scattering amplitude is
zero if the scattering angle $\theta$ is equal to $\pi /2$.
\cite{shankar,gasiorowicz}
This is because of the antisymmetry of the wave function of two identical
fermions. If the scattering amplitude for one permutation of the initial
particles going to the final particles is given by $f (\theta)$, then the
amplitude for the other permutation is given by $- f (\pi - \theta)$. The
total amplitude given by $f (\theta) - f (\pi - \theta)$ vanishes for $\theta
= \pi /2$, no matter what the form of the function $f (\theta)$ is. For two
identical anyons (the precise definition of anyons will be given in Sec.
\ref{anyons}), it is known that the scattering amplitude is not zero for
any value of $\theta$, except of course for the special case in which the
anyons are fermions. \cite{law2}

The above statements are usually made in the context of asymptotic states
where the initial and final particles are very far away from each other.
However, we can use the two-particle kernel to study what happens
when the particle positions are not far from each other.
Let $K_2 ({\vec r}_{1f}, {\vec r}_{2f}, {\vec r}_{1i}, {\vec r}_{2i}; t)$
denote the amplitude for two identical particles going from the initial
positions ${\vec r}_{1i}, {\vec r}_{2i}$ to the final positions
${\vec r}_{1f}, {\vec r}_{2f}$ in time $t$. For non-interacting particles,
$K_2$ can be written in terms of the one-particle kernel $K_1$. Namely,
\bea & & K_2 ({\vec r}_{1f}, {\vec r}_{2f}; {\vec r}_{1i}, {\vec r}_{2i}; t)
\non \\
& & = ~K_1 ({\vec r}_{1f}; {\vec r}_{1i}; t) ~K_1 ({\vec r}_{2f},
{\vec r}_{2i}; t) \non \\
& & ~~~\mp ~K_1 ({\vec r}_{2f}; {\vec r}_{1i}; t) ~K_1 ({\vec r}_{1f},
{\vec r}_{2i}; t), \label{k2} \eea
where the upper and lower signs are for fermions and bosons respectively;
in a second quantized formalism, Eq. (\ref{k2}) follows from Wick's theorem.
\cite{fetter} Using Eq. (\ref{k1}), we find that the two-particle kernel is
zero for two fermions for {\it arbitrary} values of $t$ if and only if
\beq ({\vec r}_{1i} - {\vec r}_{2i}) \cdot ({\vec r}_{1f} - {\vec r}_{2f}) = 0,
\label{cond1} \eeq
namely, the initial and final relative positions of the two particles are
perpendicular to each other. For bosons, there is no choice of initial and
final positions for which the two-particle kernel is zero for arbitrary
values of $t$.

The situation changes in an interesting way in the presence of a magnetic
field if the particles are charged. Substituting Eq. (\ref{k1mag1}) in Eq.
(\ref{k2}), we find that the two-particle kernel vanishes only if Eq.
(\ref{cond1}) is satisfied, {\it and}
\bea & & \frac{eB}{2 \hbar c} ~{\hat z} \cdot ({\vec r}_{1i} - {\vec r}_{2i})
\times ({\vec r}_{1f} - {\vec r}_{2f}) \non \\
& & = ~2 n \pi \quad {\rm for ~~fermions}, \non \\
& & = ~(2 n + 1) \pi \quad {\rm for ~~bosons}, \label{cond2} \eea
where $n= 0, \pm 1, \pm 2, \cdots$.

It is instructive to understand the conditions in Eqs.
(\ref{cond1}-\ref{cond2}) from a path integral point of view. Consider
some configurations of initial and final positions shown in Fig. 1.
The two-particle kernel in Eq. (\ref{k2}) is a
sum over all paths belonging to one of two types: paths of type $I$
are those in which one particle goes from ${\vec r}_{1i}$ to ${\vec r}_{1f}$
and the other goes from ${\vec r}_{2i}$ to ${\vec r}_{2f}$, while paths of
type $II$ are those in which one particle goes from ${\vec r}_{1i}$ to
${\vec r}_{2f}$ and the other goes from ${\vec r}_{2i}$ to ${\vec r}_{1f}$.
We see from Eq. (\ref{k1mag5}) that the {\it magnitudes} of the kernels along
paths of types $I$ and $II$ are given by
\bea & & \exp ~[ -\frac{1}{4} ~(|z_{1f} - z_{1i}|^2 ~+~ |z_{2f} - z_{2i}|^2)]
\non \\
{\rm and} & & \exp ~[ -\frac{1}{4} ~(|z_{1f} - z_{2i}|^2 ~-~ |z_{2f} -
z_{1i}|^2)] \label{mag} \eea
respectively. If these two magnitudes are not equal, then the corresponding
kernels cannot add up to zero no matter what their phases are. The condition
that the magnitudes in Eq. (\ref{mag}) should be equal implies that $(z_{1f}
- z_{2f}) (z_{1i}^* - z_{2i}^*)$ is imaginary; this is equivalent to the
perpendicularity condition in Eq. (\ref{cond1}). Next, we look at the phases
of the kernels corresponding to paths $I$ and $II$. The difference of the
phases of the two kernels is described by Eq. (\ref{cond2}) as follows. We
observe that
\beq \frac{1}{2} ~{\hat z} \cdot ({\vec r}_{1i} \times {\vec r}_{1f} ~+~
{\vec r}_{1f} \times {\vec r}_{2i} ~+~ {\vec r}_{2i} \times {\vec r}_{2f} ~
+~ {\vec r}_{2f} \times {\vec r}_{1i}) \eeq
is the area enclosed by a closed loop formed by moving in straight lines
from ${\vec r}_{1i}$ to ${\vec r}_{1f}$ to ${\vec r}_{2i}$ to ${\vec r}_{2f}$
and back to ${\vec r}_{1i}$; the area is taken to be positive for an
anticlockwise loop. The left hand side of Eq. (\ref{cond2}) can then be
interpreted as the Aharonov-Bohm phase for a particle going around the above
loop. The total phase difference between paths of types $I$ and $II$ is given
by the sum of the Aharonov-Bohm phase in Eq. (\ref{cond2}), and a phase due
to an anticlockwise exchange of the two particles which is given by $\pi$ and
$0$ for fermions and bosons respectively. (For fermions and bosons, the
exchange phase modulo $2\pi$ is the same for clockwise and anticlockwise
exchanges, but for anyons the two phases differ). The two-particle kernel
vanishes if the phases of paths $I$ and $II$ differ by $\pi$. Eq.
(\ref{cond2}) then follows in a straightforward way.

\begin{figure}[t]
\ig[width=3in]{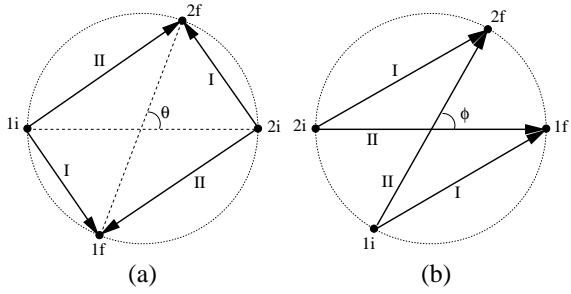}
\caption[]{Two representative configurations for anyons starting at points
$\vec{r}_{1i}$ and $\vec{r}_{2i}$ to end at points $\vec{r}_{1f}$ and
$\vec{r}_{2f}$. As the particles are indistinguishable, it is not possible to
determine which of two possible paths $I$ and $II$ each particle takes.}
\vspace{-0.1cm} \label{fig:paths}
\end{figure}

To summarize, the two-particle kernel in a magnetic field vanishes if the
initial and final relative positions are mutually perpendicular, and if the
area of the loop takes a discrete set of values which depends on the
statistics of the particles. This may be called a two-particle Aharonov-Bohm
effect \cite{buttiker1,buttiker2}; such an effect has been observed recently 
for two fermions \cite{neder}.

Looking at Eqs. (\ref{k1mag1}) and (\ref{k1mag3}), we see that
the condition for the vanishing of the two-particle kernel is
given by Eqs. (\ref{cond1}) and (\ref{cond2}), whether one considers all the
states in a magnetic field or only the states in the LLL. For the general
case of anyons, we will find the condition for the vanishing of the
two-particle kernel in the LLL in the next section; the corresponding
condition for all states in a magnetic field is not yet known for anyons.

It is useful to derive the two-particle kernel for fermions and bosons
in a different way, using center of mass and relative coordinates; this
can then be generalized to find the kernel for two anyons. We define the
center of mass coordinate ${\vec R} = ({\vec r}_1 + {\vec r}_2)/2$, momentum
${\vec P} = {\vec p}_1 + {\vec p}_2$, mass $2\mu$ and charge $Q = 2e$, and
the relative coordinate ${\vec r} = {\vec r}_1 - {\vec r}_2$, momentum
${\vec p} = ({\vec p}_1 - {\vec p}_2)/2$, mass $\mu /2$ and charge $q =
e/2$. The Hamiltonian then decouples as
\bea H &=& \frac{1}{4\mu} ~(P_x + \frac{QB}{2c} Y)^2 ~+~ \frac{1}{4\mu} ~
(P_y - \frac{QB}{2c} X)^2 \non \\
& & +~ \frac{1}{\mu} ~(p_x + \frac{qB}{2c} y)^2 ~+~ \frac{1}{\mu} ~(p_y -
\frac{qB}{2c} x)^2. \label{ham3} \eea
The eigenstates of this Hamiltonian are given by products of
eigenstates of the center of mass and relative coordinate systems, and
the energies are given by sums of the individual energies. In
the LLL, the eigenstates of the centre of mass system are given by
\beq \psi_n ({\vec R}) = \sqrt{\frac{QB}{n! ~2\pi \hbar c}} \left( Z
\sqrt{\frac{QB}{2\hbar c}} \right)^n ~\exp [ -\frac{QB}{4\hbar c} ~|Z|^2], \eeq
where $Z=X+iY$, and $n=0,1,2,\cdots$. The eigenstates of the relative system
are given by
\beq \psi_p ({\vec r}) = \sqrt{\frac{qB}{p! ~2\pi \hbar c}} \left( z
\sqrt{\frac{qB}{2\hbar c}} \right)^p ~\exp [ -\frac{qB}{4\hbar c} ~|z|^2], \eeq
where $p = 0,2,4,\cdots$ for bosons and $1,3,5,\cdots$ for fermions. The
condition on $p$ arises from the fact that the wave functions must be
symmetric (antisymmetric) under the exchange of the two particles,
$z \to - z$, for bosons and fermions respectively.

The two-particle kernel in imaginary time is given by
\bea & & K_2 ({\vec r}_{1f}, {\vec r}_{2f}; {\vec r}_{1i}, {\vec r}_{2i};
\tau) \non \\
& & =~ \sum_n ~\psi_n ({\vec R}_f) ~\psi_n^* ({\vec R}_i) ~e^{-E_n \tau /
\hbar} \non \\
& & ~~~~\times ~\sum_p ~\psi_p ({\vec r}_f) ~\psi_p^* ({\vec r}_i) ~e^{-E_p
\tau /\hbar}. \label{k2imag} \eea
Absorbing the Landau length in the length scale and shifting the ground state
energy to zero as before, we find that the kernel in the LLL is given by
\bea & & K_2 ({\vec r}_{1f}, {\vec r}_{2f}; {\vec r}_{1i}, {\vec r}_{2i})
\non \\
& & =~ \frac{1}{(2\pi l^2)^2} ~\exp ~[-\frac{1}{4} (|z_{1f}|^2 + |z_{2f}|^2 +
|z_{1i}|^2 + |z_{2i}|^2) \non \\
& & ~~~~~~~~~~~~~~~~~~~~+ \frac{1}{4} (z_{1f} + z_{2f})(z_{1i}^* + z_{2i}^*)]
\non \\
& & \times ~\sinh [\frac{1}{4} (z_{1f} - z_{2f}) (z_{1i}^* - z_{2i}^*)] \quad
{\rm for ~~ fermions}, \non \\
& & \times ~\cosh [\frac{1}{4} (z_{1f} - z_{2f}) (z_{1i}^* - z_{2i}^*)] \quad
{\rm for ~~ bosons}. \label{k2fermibose} \eea
One can check that Eq. (\ref{k2fermibose}) agrees with the results obtained
by combining Eqs. (\ref{k1mag5}) and (\ref{k2}).

The kernel in Eq. (\ref{k2fermibose}) satisfies the reproducing property
\bea \psi ({\vec r}_{1f}, {\vec r}_{2f}) &=& \int ~d^2{\vec r}_{1i} ~d^2
{\vec r}_{2i} ~K_2 ({\vec r}_{1f}, {\vec r}_{2f}; {\vec r}_{1i}, {\vec r}_{2i})
\non \\
& & ~~~~\times ~\psi ({\vec r}_{1i}, {\vec r}_{2i}). \label{k2reproduce} \eea
This can be shown by noting that an arbitrary antisymmetric (symmetric)
two-particle wave function $\psi$ in the LLL can be written as
\bea \psi ({\vec r}_1, {\vec r}_2) &=& \sum_{n,p=0}^\infty ~c_{n,p} ~ (z_1 +
z_2)^n ~(z_1 - z_2)^p \non \\
& & ~~~~\times ~\exp [ -\frac{1}{4} (|z_1|^2 + |z_2|^2)], \eea
where $p$ is restricted to odd (even) integers for fermions (bosons)
respectively, and then proving that the reproducing property holds for
each term $(z_1 + z_2)^n ~(z_1 - z_2)^p$ separately.

\subsection{Anyons in the lowest Landau level}
\label{anyons}

We will now derive the two-particle kernel for anyons in the LLL. Anyons are
particles living in two dimensions with the property that when two of them
are interchanged in an anticlockwise (clockwise) direction, the wave function
picks up a phase of $e^{i\pi \al}$ ($e^{-i\pi \al}$) respectively; here $\al$
is a real parameter lying in the range $-1 < \al \le 1$.
\cite{leinaas,wilczek1} (Bosons and fermions correspond to the special cases
$\al = 0$ and 1 respectively). For instance, for a system of two particles,
we demand that if we continuously vary the arguments ${\vec r}_1$ and
${\vec r}_2$ of a wave function so as to interchange them in an anticlockwise
sense (without ever making them equal to each other), the final wave function
must be $e^{i\pi \al}$ times the initial wave function. These statements are
true in a formalism in which the Hamiltonian does not explicitly depend on
$\al$, but the wave functions are multi-valued. There is an alternative
formalism in which the Hamiltonian depends on $\al$, and the wave functions
are single-valued. A brief discussion of these two approaches is given in
Appendix B. Further details can be found in Refs.
\onlinecite{wilczek2,forte,lerda,khare,myrheim}.

In a fractional quantum Hall system with filling fraction given by $1/m$,
where $m$ is an odd integer, the quasi-particles are believed to behave
as anyons. The charge of a quasi-hole (quasi-electron) is $-e/m$ ($e/m$),
where $e$ is the charge of an electron, and these are anyons with $\al =
1/m~$ ($-1/m$) respectively. \cite{laughlin2,halperin,arovas,myrheim} We will
henceforth discuss the case of quasi-holes. (Appendix C provides a brief
derivation of the charge and statistics of quasi-holes based on the Laughlin
wave functions). For simplicity, we will ignore the Coulomb interaction
between two quasi-holes; our interest is in quasi-hole propagation over
short distances of the order of $300 ~nm$ \cite{picciotto,saminadayar}, while
Coulomb interaction effects only become significant over longer distances
\cite{law2}. (These effects can, in principle, be treated perturbatively
\cite{kivelson}.) We will therefore take quasi-holes as being described by
a model of anyons in the LLL with $\al = 1/m$.

For a two-particle system in the LLL, it is simple to impose the exchange phase
condition on the wave function since it only affects the relative coordinates
$(z,z^*)$. Since the Hamiltonian takes the form given in Eq. (\ref{ham2}), up
to some numerical factors, we see that a wave function of the form
$z^{2p+\al}$ times a Gaussian will do the job, where $p = 0,1,2,\cdots$.
The normalized eigenstates of the relative system are therefore given by
\bea \psi_p ({\vec r}) &=& \sqrt{\frac{qB}{\Gamma (2p + \al + 1) ~2\pi
\hbar c}} ~\left( z \sqrt{\frac{qB}{2\hbar c}} \right)^{2p+\al} \non \\
& & \times ~\exp ~[ -\frac{qB}{4\hbar c} ~|z|^2], \label{psip} \eea
where $q$ is half the charge of an anyon.

Absorbing the Landau length $\sqrt{{\hbar c}/eB}$ in the coordinates $z$
and $z^*$ in Eq. (\ref{psip}), we obtain the two-anyon kernel in the LLL,
\bea & & K_2 ({\vec r}_{1f}, {\vec r}_{2f}; {\vec r}_{1i}, {\vec r}_{2i})
\non \\
& & = \frac{1}{(2\pi m l^2)^2} \exp [-\frac{1}{4m} (|z_{1f}|^2 + |z_{2f}|^2
+ |z_{1i}|^2 + |z_{2i}|^2) \non \\
& & ~~~~~~~~~~~~~~~~~~~~~+ \frac{1}{4m} (z_{1f} + z_{2f}) (z_{1i}^* +
z_{2i}^*)] \non \\
& & \times ~\sum_{p=0}^\infty ~\frac{[(z_{1f} - z_{2f})(z_{1i}^* -
z_{2i}^*)/4m]^{2p+1/m}}{\Gamma (2p + 1/m +1)}. \label{k2anyon} \eea
We will call this the Laughlin kernel since it seems to have been first
written down by him. \cite{laughlin2} (The factors of $1/m$ multiplying
$zz^*$ appear because the anyon has a charge $-e/m$ but the Landau
length rescaling does not involve $m$). Eq. (\ref{k2anyon}) satisfies the
reproducing property in Eq. (\ref{k2reproduce}) if $\psi$ has the LLL form
\bea \psi ({\vec r}_1, {\vec r}_2) &=& \sum_{n,p=0}^\infty ~c_{n,p} ~ (z_1 +
z_2)^n ~(z_1 - z_2)^{2p+1/m} \non \\
& & ~~~~\times ~\exp [ -\frac{1}{4m} (|z_1|^2 + |z_2|^2)]. \eea

For any finite value of $m$, the Laughlin kernel trivially vanishes if
$z_{1i} = z_{2i}$ or $z_{1f} = z_{2f}$. We will now study the non-trivial
zeros of the kernel, in particular of the function $z^{1/m} F_m (z)$, where
\bea z &=& \frac{1}{4m} ~(z_{1f} - z_{2f})(z_{1i}^* - z_{2i}^*), \non \\
{\rm and} ~~ F_m (z) &=& \sum_{p=0}^\infty ~
\frac{z^{2p}}{\Gamma (2p + 1/m +1)}. \label{fmz} \eea
For $m=1$ (fermions), we see that $z F_1 (z) = \sinh z$ vanishes when $z = i
n \pi$, where $n$ is an integer; the fact that $z$ is imaginary is equivalent
to Eq. (\ref{cond1}), while the fact that $z = i n \pi$ is equivalent to the
condition given in Eq. (\ref{cond2}) for fermions. In general, the function
in Eq. (\ref{fmz}) is related to confluent hypergeometric functions
\cite{abramowitz} as
\bea & & 2 ~\Gamma (1+1/m) ~F_m (z) \non \\
& & = ~M(1, 1+1/m, z) ~+~ M(1, 1+1/m, -z). \eea
Let us look for zeros of $F_m (z)$ at the values $z=iy$, where $y$ is real
and positive. For $y \to \infty$, the asymptotic behavior of confluent
hypergeometric functions \cite{abramowitz} imply that, for $m > 1/2$,
\beq 2 ~\Gamma (1+1/m) ~F_m (iy) ~\simeq ~\frac{2}{y^{1/m}} ~\cos ~(y -
\frac{\pi}{2m}). \label{asymp} \eeq
This has zeros at the points
\beq y_n ~=~ (n ~-~ \frac{1}{2} ~+~ \frac{1}{2m}) ~\pi, \label{zeros} \eeq
where $n=1,2,3, \cdots$. Since this result is based on asymptotic formulae,
we expect it to be valid only for $n \to \infty$. However, Eq. (\ref{zeros})
is exact for the two special cases $m=1$ and $\infty$, since
\beq F_1 (iy) ~=~ \frac{\sin y}{y}, \quad {\rm and} \quad F_\infty (iy) ~=~
\cos y. \label{fermibose} \eeq
Numerically, we find that Eq. (\ref{zeros}) is quite accurate even for small
values of $n$, for any value of $m \ge 1$. Using the results in Ref.
\onlinecite{abramowitz}, we can obtain a more precise formula which shows
how the result in (\ref{zeros}) is approached for large $n$, namely,
\bea y_n &=& (n ~-~ \frac{1}{2} ~+~ \frac{1}{2m}) ~\pi \non \\
& & +~ \frac{(-1)^n ~\sin (\pi/m) ~\Gamma (2-1/m)}{\pi ~[\pi
(n-1/2+1/(2m))]^{2-1/m}}. \label{zeros2} \eea
A comparison between the results obtained from Eqs. (\ref{zeros}),
(\ref{zeros2}) and numerical calculations is presented in Table 1 for the
first four values of $y_n /\pi$ for $m=3$.

\begin{table}[h] \begin{center} \begin{tabular}{|c|c|c|c|} \hline
$n$ & Eq. (\ref{zeros}) & Eq. (\ref{zeros2}) & Numerical \\ \hline
1 & 0.6667 & 0.6436 & 0.6509 \\
2 & 1.6667 & 1.6717 & 1.6711 \\
3 & 2.6667 & 2.6644 & 2.6645 \\
4 & 3.6667 & 3.6680 & 3.6680 \\ \hline \end{tabular} \end{center}
\caption{Values of $y_n /\pi$ for the first four zeros of the Laughlin kernel
for $m=3$, obtained from Eqs. (\ref{zeros}), (\ref{zeros2}) and numerical
calculations respectively.}
\end{table}

Finally, we note that if $z=iy_n$ is a zero of the function $F_m (z)$, the
symmetry of that function implies that $z=-iy_n$ is also a zero. We conclude
that the Laughlin kernel has zeros at $z=\pm iy_n$, where $y_n$ is
approximately given by Eq. (\ref{zeros}).

We can understand the result in Eq. (\ref{zeros}) using path integral
arguments similar to the ones presented after Eq. (\ref{cond2}). The Laughlin
kernel can be thought of as the sum over all paths of types $I$ and $II$ as
described earlier. We see from Eq. (\ref{k1mag5}), with the charge being set
equal to $e/m$, that the magnitudes of the kernels along paths $I$ and $II$
are given by
\bea & & \exp ~[\frac{1}{4m} ~(|z_{1f} - z_{1i}|^2 + |z_{2f} - z_{2i}|^2)]
\non \\
{\rm and} & & \exp ~[\frac{1}{4m} ~(|z_{1f} - z_{2i}|^2 + |z_{2f} - z_{1i}|^2)]
\eea
respectively. The condition that the magnitudes should be equal is equivalent
to the statement that $z$ is imaginary. Next, we look at the phases
corresponding to kernels of paths $I$ and $II$. The difference of the phases
of the two kernels is given by the sum of an Aharonov-Bohm phase which turns
out to be $-i(z_{1f} - z_{2f})(z_{1i}^* - z_{2i}^*)/(2m) = -i2z = 2y$ (we
recall that the anyon charge is $-e/m$), and
an exchange phase which depends on the sign of $y$ as follows. For $y > 0$
and $< 0$, the path obtained by going from $z_{1i}$ to $z_{1f}$ to $z_{2i}$
to $z_{2f}$ and back to $z_{1i}$ forms a closed loop in the anticlockwise and
clockwise directions respectively; the exchange phase is therefore given by
$-(\pi /m) sgn (y)$, where $sgn (y) \equiv 1$ if $y > 0$ and $-1$ if $y < 0$
(see Appendix B). Hence, the kernels of paths $I$ and $II$ will add up to zero
if $2y - (\pi/m) sgn (y)$ is an odd multiple of $\pi$. This agrees with
the equation for the zeros at $z=\pm i y_n$ given in Eq. (\ref{zeros}).

A consequence of the above analysis is that if the initial and final positions
of the two anyons lie on a circle of radius $r$, with $z_{2i} = - z_{1i} = r$
and $z_{2f} = - z_{1f} = r e^{i\theta}$, as shown in Fig. 1 (a), then the
Laughlin kernel vanishes if the scattering angle $\theta$ is $\pi/2$, and $r$
takes a discrete set of values given by
\beq 2r^2 ~=~ \frac{2 \pi \hbar c}{eB} ~m~ (n ~-~ \frac{1}{2} ~+~
\frac{1}{2m}), \label{area1} \eeq
where we have restored the Landau length. Namely, the kernel vanishes if the
rectangle in Fig. 1 (a) is a square, and its area $2 r^2$ is quantized as
given above.

We can consider a different geometry in which the initial and final positions
of the anyons again lie on a circle, but the angle between the initial
positions is $\phi$, as is the angle between the final positions. For
instance, we can have $z_{1f} = r = - z_{2i}$ and $z_{2f} = r e^{i\phi} = -
z_{1i}$, as shown in Fig. 1 (b). As $r$ or $\phi \to 0$, we see that the
Laughlin kernel in Eq. (\ref{k2anyon}) goes to zero with a power law, $[r
\sin (\phi / 2)]^{2/m}$. This is related to the fact that anyons obey a
generalized exclusion principle; this is discussed in Appendix D.

Figures 2 (a) and (b) show the magnitude of the two-particle kernel as
functions of the radius $r$ and the angles $\theta$ and $\phi$ respectively
as discussed above, for the case $m=3$. Figure 2 (a) shows some zeros of the
kernel which lie on the line $\theta = \pi /2$, while Fig. 2 (b) indicates
that the kernel vanishes as $r$ or $\phi$ approaches zero.

\begin{figure}[htb]
\ig[width=1.35in]{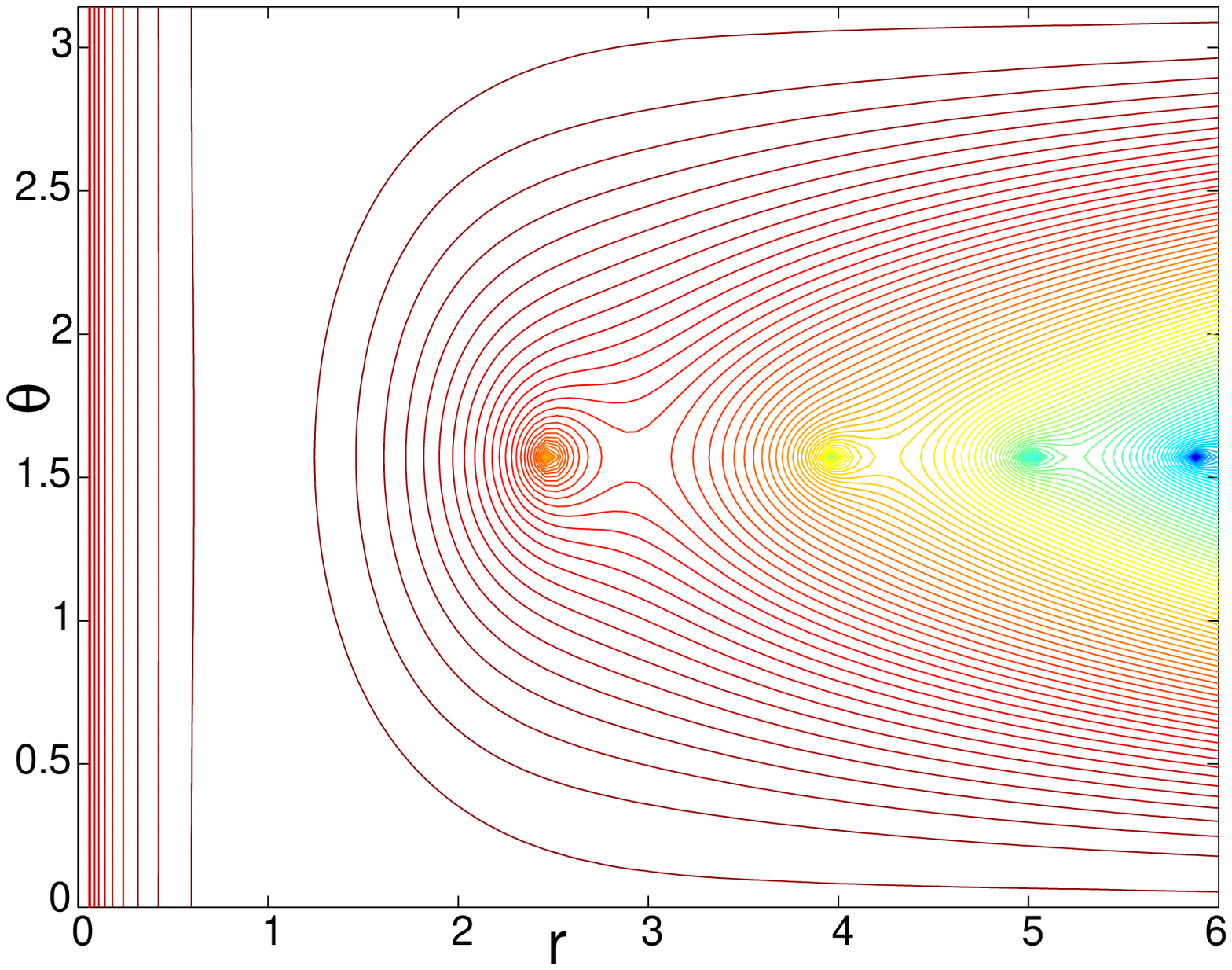}\hspace{.6cm}\ig[width=1.4in]{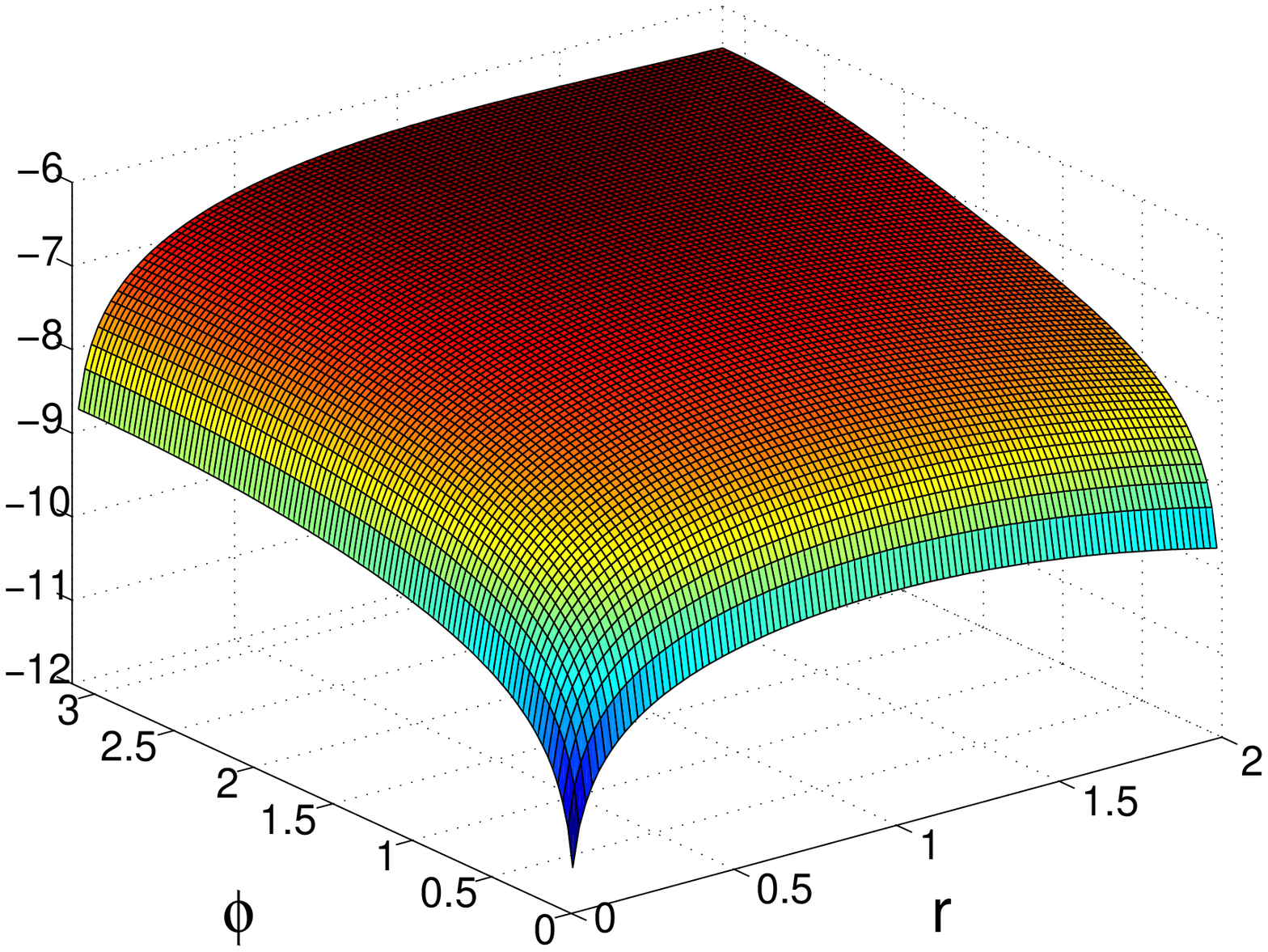}
\vspace{0.5cm}
\leftline{~~~~~~~~~~~~~~~~~~~~(a)~~~~~~~~~~~~~~~~~~~~~~~~~~~~~~~(b)}
\vspace{-1cm}
\caption[]{(Color online) Magnitude of the two-anyon kernel versus the
radius $r$ and the angle $\theta$ and $\phi$ for the configurations shown
in Figs. 1 (a) and (b) respectively, for $m=3$. Fig. 2 (a) shows the first
four zeros of the kernel lying on the line $\theta = \pi /2$, while Fig. 2
(b) shows that the kernel vanishes as either $r$ or $\phi$ approaches zero.}
\vspace{-0.1cm} \label{kernel}
\end{figure}

Eq. (\ref{area1}) can be generalized to the case of filling fractions different
from $1/m$ where there are incompressible fractional quantum Hall states. In
general, the ratio of the quasi-particle charge $e^*$ to the electron charge 
$e$ and the fractional statistics parameter $\al$ need not be equal to each 
other as they are for the Laughlin states. The two-particle kernel in the 
general case is given by
\bea & & K_2 ({\vec r}_{1f}, {\vec r}_{2f}; {\vec r}_{1i}, {\vec r}_{2i}) 
\non \\
& & = \left( \frac{e^*}{2\pi e l^2} \right)^2 \exp [-\frac{e^*}{4e}
(|z_{1f}|^2 + |z_{2f}|^2 + |z_{1i}|^2 + |z_{2i}|^2) \non \\
& & ~~~~~~~~~~~~~~~~~~~~~~+ \frac{e^*}{4e} (z_{1f} + z_{2f}) (z_{1i}^* +
z_{2i}^*)] \non \\
& & \times ~\sum_{p=0}^\infty ~\frac{[(z_{1f} - z_{2f})(z_{1i}^* -
z_{2i}^*)/4m]^{2p+\al}}{\Gamma (2p + \al +1)}. \label{general} \eea
One can then show, using arguments similar to the ones which led from Eq. 
(\ref{k2anyon}) to (\ref{area1}), that the two-particle kernel will have 
zeros when the rectangle in Fig. 1 (a) is a square with an area given by
\beq 2r^2 ~=~ \frac{2 \pi \hbar c}{eB} ~\frac{e}{e^*} ~(n ~-~ \frac{1}{2} ~+~
\frac{\al}{2}). \label{area2} \eeq

An example of the general case presented in Eq. (\ref{general}) is provided by
quasi-particles of the Jain states; these states appear at filling fractions of
the form $\nu = p/(2np \pm 1)$, where $p$ and $n$ are integers. \cite{jain1}
For $p>1$, the ratio $e^*/e$ is not equal to $\al$, unlike the special case 
of the Laughlin states which correspond to $p=1$. For $\nu = p/(2np + 1)$, 
the fractional charge of the Jain state quasi-particles is given by $e^*= 
e/(2np+1)$. In one formulation, the statistical parameter has the form $\al 
= 1-2n/(2np+1)$; this renders the $p>1$ Jain states more fermion-like in 
that $\al > 1/2$, in comparison with the Laughlin states which have $\al < 
1/2$. \cite{fradkin} 

However, it is likely that a description of the quasi-particles for the Jain
states will require us to beyond the LLL. One way to see this is to consider 
the composite fermion theory for fractional quantum Hall states. \cite{jain1}
For instance, the state with $\nu = p/(2np +1)$ corresponds to composite 
fermions filling $p$ Landau levels. We will therefore discuss in Sec. III. D 
how the effects of higher Landau levels may be computed.

\subsection{Experimental Realization}

Based on the above analysis, we propose an experiment for measuring the charge
and statistics of the quasi-particles. Consider a configuration in which there
are two sources and two detectors of anyons, corresponding to the positions
$z_{1i}, z_{2i}$ and $z_{1f}, z_{2f}$ respectively, which are arranged as a
square as shown in Fig. \ref{fig:paths}. a ($\theta = \pi/2$). The sources and
detectors can be created by bringing four edge states close together via a gate
potential and setting the edges identified as the sources at a higher potential
than those identified as the detectors, thus generalizing the principles used
for observing the propagation of a single quasi-particle using a single source
and a single detector. \cite{picciotto,saminadayar} Holding the area of the
configuration and the filling fraction in the bulk fixed, one can gradually
change the magnetic field and determine some successive values of the field
where a simultaneous observation of the anyon tunneling current in the two
detectors gives a null result. According to Eq. (\ref{area2}), a plot of the
magnetic field $B$ versus the number of the zero $n$ will be a straight line
whose slope is proportional to the charge $e^*$ and whose intercept gives the
value of $(\al - 1)/2$. Even if one mis-identifies the values of $n$ by a
constant integer, the intercept will correctly give the fractional part of
$(\al - 1)/2$; this will be sufficient to identify the value of $\al$ in the
range $[-1,1]$.

Experimentally, there may be a small uncertainty in the precise locations of
the sources and detectors; further, the edge confining potential may be 
somewhat soft, with a width of the order of several magnetic lengths. 
Both of these would lead to an uncertainty in the area $2r^2$ in Eq.
(\ref{area2}). This would lead to some uncertainty in the value of $e^*$
obtained from the slope of the straight line, but there would still be no
uncertainty in the value of $\al$ obtained from the intercept. We therefore
expect a determination of the statistics parameter via this method to be more
robust than the charge.

In the discussion above, we have proposed holding the area of a certain
configuration and the filling fraction in the bulk fixed, and finding some
successive values of the magnetic field where the two-particle kernel vanishes.
This is a reasonable procedure because as long as we are at a quantum Hall
plateau, varying the magnetic field within some range does not change the
filling fraction. However, this argument does require the region within the
area of interest to be sufficiently clean and free of pinning impurities, so
that no additional anyons appear or disappear there as the magnetic field is
being varied. If additional anyons get trapped in that region, they would
contribute to the two-particle kernel and destroy the simple relation between
the magnetic field and the zeros of the kernel. Furthermore, our arguments do
not take into account the invasive nature of the gates and the effects of the
quantum Hall boundary. While we consider the effect of a boundary in the next
section, a full-fledged treatment of the geometry proposed here is beyond the
scope of the paper. However, we expect that with all considerations taken into
account, the system will still exhibit effects of fractional charge and
fractional statistics in coincidence measurements.

It has been argued that fractional statistics is a valid concept only if the
distance between two quasi-particles is greater than about $10$ times the
Landau length $l$. \cite{jain1} In typical experiments involving tunneling
between quantum Hall edges, the tunneling distance is about $300 ~nm$.
\cite{picciotto,saminadayar} This is much larger than $l$ which is of the order
of $10 ~nm$ for a filling fraction of $1/3$. [That the tunneling amplitude is 
not negligible even though the ratio of the tunneling distance to the magnetic
length is about 30 is probably due to the softness of the edge confining
potential which gives it a width of several magnetic lengths. The implication
of this softness for the determination of the charge and statistics parameter
has been discussed above.] In terms of the configuration in
Fig. 1 (a), we have $r/l \sim 30$. While such a large ratio certainly justifies
the use of fractional statistics, it gives the value of the integer $n$ in Eq.
(\ref{area2}) to be about $10^3$; it would be difficult to distinguish a
fraction from such large values of $n$. To reduce the value of $n$ as far as
possible and at the same time ensure that the idea of fractional statistics
remains valid, we require values of $r/l$ which are not much larger than 10.

\subsection{Higher Landau levels}

There are several instances where a projection to the LLL will not suffice and
higher Landau levels need to be invoked. Examples include finite temperature,
length scales shorter than the magnetic length and time scales faster than the
inverse
cyclotron frequency, quasi-particles in non-Laughlin states such as most of the
Jain states, and tunneling of quasi-particles in certain situations. Moreover,
it is only by including higher Landau levels that one can consider dynamics; 
as we saw earlier, the kernel in the LLL has a trivial dependence on time.
In what follows, as a start, we will calculate the effect of higher Landau
levels on the two-anyon kernel using an imaginary time formalism. As before, 
we will ignore all interactions between the two particles apart from the 
exchange statistics. [An imaginary time formalism is directly relevant to a 
calculation of the density matrix at finite temperature; the imaginary time 
$\tau$ is inversely related to the temperature as $\tau = \hbar /(k_B T)$]. 

If the contribution of all Landau levels is taken into account, we expect the 
kernel to depend on the time coordinate. However, for the special cases of 
fermions and bosons, it turns out that the locations of the zeros of the 
kernel are independent of time and remain exactly at the same points where 
they were in the LLL calculation. This can be seen by substituting the 
expression in Eq. (\ref{k1mag2}) in Eq. (\ref{k2}); we then find that the 
condition for the vanishing of the two-particle kernel is given by Eqs. 
(\ref{cond1}-\ref{cond2}) precisely as in the LLL.

We have not been able to find an analytic expression for the 
two-particle kernel for anyons in an arbitrary magnetic field analogous to 
the LLL expression given in Eq. (\ref{k2anyon}). However, we can make some 
general statements about the wave functions and energy levels \cite{myrheim},
and then about the zeros of the kernel in a particular limit of the imaginary
time $\tau$. The two-particle kernel can again be factorized into a product of
centre of mass and relative coordinate kernels as shown in Eq. (\ref{k2imag}).
The centre of mass kernel is given by the expression in Eq. (\ref{k1mag2}), 
with the charge, cyclotron frequency and coordinates in that equation being 
replaced by the appropriate centre of mass quantities; it is clear that this 
kernel does not have any zeros.
Only the relative coordinate kernel is sensitive to the statistics
parameter, and only this kernel can possibly have zeros. We will therefore
consider only the relative coordinate kernel henceforth. Absorbing the 
appropriate Landau length in the complex coordinate $z = z_1 - z_2$, and 
assuming as in Eq. (\ref{ham2}) that the relative coordinate wave functions 
are of the form $\psi = f(z,z^*) \exp (- |z|^2/4)$, we obtain the eigenvalue 
equation ${\tilde H} f = E f$, where
\beq {\tilde H} ~=~ \hbar \Omc ~[- ~2 \frac{\partial}{\partial z}
\frac{\partial}{\partial z^*} ~+~ z^* \frac{\partial}{\partial z^*}],
\label{ham5} \eeq
where $\Omc$ is the cyclotron frequency for the relative coordinate system (it
is determined by the charge $e^*$ and the reduced mass), and we have shifted
the ground state energy to zero. Let us assume that the statistics parameter
$\al$ lies in the range $0 \le \al \le 1$. We then find that there are two
types of eigenfunctions of Eq. (\ref{ham5}):

\noi (i) $f_{n,N} = z^{2n + \al}$ multiplied by a polynomial of degree $N$ in
$zz^*$, where $n = 0,1,2, \cdots$, and

\noi (ii) $f_{-n,N} = z^{*2n - \al}$ multiplied by a finite polynomial of
degree $N$ in $zz^*$, where $n = 1,2, \cdots$.

\noi The allowed values of $n$ are determined by the requirement that the wave
function should not diverge as $z \to 0$. The form of the Hamiltonian in Eq.
(\ref{ham5}) implies that the eigenfunctions of types (i) and (ii) have
energies $N \hbar \Omc$ and $(N + 2n - \al) \hbar \Omc$ respectively. The
LLL wave functions are a special case of the type (i) states, with $N=0$.
[Note that for $\al = 0 (1)$, the wave functions given in (i) and (ii) are
even (odd) under $z \to - z$, as expected from the symmetry (antisymmetry)
of bosonic (fermionic) wave functions under an exchange of two particles.
It is also interesting to observe that the energy levels of types (i) and 
(ii) do not coincide with each other if $\al$ is not equal to 0 or 1].

We can get an idea of the effect of higher Landau levels on the two-particle
kernel as follows. For $0 \le \al \le 1$, the lowest two sets of excited states
are given by type (i) states with any $n$ but $N=1$ (these have $E=\hbar
\Omc$), and the type (ii) state with $n=1$ and $N=0$ (this has $E=(2-\al) 
\hbar \Omc$). Let us define $\xi = e^{-\Omc \tau}$ and assume that this is 
small. We will ignore contributions from states with energies $E \ge 2
\hbar \Omc$; hence we will keep terms only up to $\xi$ and $\xi^{2-\al}$,
and ignore terms of order $\xi^2$ and higher. Our results will therefore be 
applicable for a time separation $\tau$ long compared to the inverse
cyclotron frequency. To this order, the contribution to the kernel comes from
the following normalized wave functions: 
\bea \psi_{n,0} ({\vec r}) &=& \frac{1}{\sqrt{2 \pi \Gamma (2n + \al + 1)}}
\left( \frac{z}{\sqrt{2}} \right)^{2n+\al} e^{-|z|^2 /4}, \non \\
\psi_{n,1} ({\vec r}) &=& \sqrt{\frac{2n + \al + 1}{2 \pi \Gamma (2n +
\al + 1)}} \left( \frac{z}{\sqrt{2}} \right)^{2n+\al} \non \\
& & \times ~\left( 1 ~-~ \frac{zz^*}{2(2n + \al + 1)} \right) e^{-|z|^2 /4},
\non \\
\psi_{-1,0} ({\vec r}) &=& \frac{1}{\sqrt{2 \pi \Gamma (3 - \al)}}
\left( \frac{z^*}{\sqrt{2}} \right)^{2-\al} e^{-|z|^2 /4}. \label{wave2} \eea
[The definition of $z$ in Eq. (\ref{wave2}) differs by a numerical factor
from the definition used in Eq. (\ref{psip})]. The contribution to the
two-particle kernel from the above states is given by
\bea & & K_2 ({\vec r}_{1f}, {\vec r}_{2f}; {\vec r}_{1i}, {\vec r}_{2i}; \tau)
\non \\
&=& \sum_{n=0}^\infty ~\psi_{n,0} ({\vec r}_{1f} - {\vec r}_{2f}) ~\psi_{n,0}^*
({\vec r}_{1i} - {\vec r}_{2i}) \non \\
&~+ & \sum_{n=0}^\infty ~\psi_{n,1} ({\vec r}_{1f} - {\vec r}_{2f}) ~
\psi_{n,1}^* ({\vec r}_{1i} - {\vec r}_{2i})~ \xi \non \\
&~+& \psi_{-1,0} ({\vec r}_{1f} - {\vec r}_{2f}) ~ \psi_{-1,0}^*
({\vec r}_{1i} - {\vec r}_{2i}) ~\xi^{2-\al}. \label{k2higher} \eea
Let us now introduce the variables $x=(z_{1f} -z_{2f})(z_{1i}^* -z_{2i}^*)/2$,
$x_i = |z_{1i} - z_{2i}|^2 /2$ and $x_f = |z_{1f} - z_{2f}|^2 /2$; note that
$x_i x_f = |x|^2$. Eq. (\ref{k2higher}) can then be written in terms of
confluent hypergeometric functions \cite{abramowitz} as
\bea K_2 &=& \frac{x^\al ~e^{-(x_i+x_f)/2}}{2\pi \Gamma (\al + 1)} ~[~
\{ 1 ~+~ (1 - x_i - x_f) ~\xi \} \non \\
& & \times ~\frac{1}{2} \{ M(1, 1+\al, x) + M(1, 1+\al, -x) \} \non \\
& & + ~(x ~+~ x^*) ~\xi \non \\
& & \times ~\frac{1}{2} \{ M(1, 1+\al, x) - M(1, 1+\al, -x) \} \non \\
& & +~ \al ~\xi] ~+~ \frac{x^{*2-\al} ~e^{-(x_i+x_f)/2}}{2\pi \Gamma
(3 - \al)} ~\xi^{2-\al}. \label{k2higher2} \eea

Let us now find where the zeros of the expression in Eq. (\ref{k2higher2})
are. Since we know that the zeros of the kernel for bosons and fermions
continue to satisfy the perpendicularity condition in Eq. (\ref{cond1}),
let us assume that this remains true for anyons in general and see if
this gives us the locations of the zeros. This condition implies that $x$ 
is imaginary in Eq. (\ref{k2higher2}); let us set $x=iy$, where $y > 0$.
Upon ignoring terms of order $\xi^2$ and higher, we find that the zeros of
(\ref{k2higher2}) are given by the relation
\beq Re ~M(1, 1+\al, iy) ~=~ - \al ~\xi ~+~ \frac{\Gamma (\al + 1)}{\Gamma
(3 - \al)} ~y^{2-2\al} ~\xi^{2-\al}, \label{k2higher3} \eeq
where $Re$ denotes the real part. [Note that the condition for the zeros
of the LLL kernel is recovered when one takes the limit $\tau \to \infty$,
i.e., sets $\xi = 0$ in Eq. (\ref{k2higher3})]. For $\al =0$, $Re ~M(1,1+\al,
iy) = \cos y$, and we find that Eq. (\ref{k2higher3}) is satisfied if $y=
(n-1/2)\pi$, where $n=1,2,3,\cdots$ (we ignore the last term in
(\ref{k2higher3}) which is of order $\xi^2$); this is the expected result for
bosons as given by Eq. (\ref{fermibose}). For $\al =1$, $Re ~M(1,1+\al,iy) =
\sin y /y$, and Eq. (\ref{k2higher3}) is satisfied if $y=n\pi$, where
$n=1,2,3,\cdots$; this is the correct result for fermions as given by Eq.
(\ref{fermibose}).

For $0 < \al < 1$, we find that Eq. (\ref{k2higher3}) is satisfied for a
discrete set of values of $y$, but these values depend on $\xi = e^{-\Omc
\tau}$ and therefore on the time $\tau$. Using the asymptotic expression given
in Eq. (\ref{asymp}) for $y \to \infty$, we find that Eq. (\ref{k2higher3})
takes the form
\beq \cos ~(y - \frac{\al \pi}{2}) ~=~ - \al y^\al ~\xi ~+~ \frac{\Gamma
(\al + 1)}{\Gamma (3 - \al)} ~y^{2-\al} ~\xi^{2-\al} . \label{k2higher4} \eeq
If $\xi$ is very small, the above equation can be used to find the correction
from the zeros of the LLL kernel which lie at $y_n = (n -1/2 + \al /2) \pi$;
the correction can be seen to go to zero exponentially as $\tau \to \infty$.
Eq. (\ref{k2higher4}) shows that the zeros of the kernel will deviate 
appreciably from their locations in the LLL if either $\tau$ becomes 
comparable to the inverse cyclotron frequency, $\Omc \tau \sim 1$, or if 
a distance scale like the radius of the circle in Fig. 1 becomes much larger 
than the magnetic length, $y \gg 1$. This gives an idea of the correlation 
length and time scales in this problem; a discussion of the length and time 
scales in two-particle problems appearing in other areas of physics is given 
in Ref. \onlinecite{baym}.

Although the above calculation was based only on the lowest few Landau levels,
we conclude that in general, the locations of the zeros of the two-anyon
kernel in an arbitrary magnetic field depend on the time. Thus the simple
geometrical understanding of the zeros of the LLL kernel (based on the phase
difference between two paths) given in Sec. \ref{anyons} needs to be modified
when one considers the effect of higher Landau levels.

While the above treatment using imaginary time may be a valid starting point 
for studying finite temperature and tunneling, other processes would require 
using real time. Examples may include the propagation of Jain state 
quasi-particles; however, an explicit effective wave function description 
such as the one we have used here for Laughlin states is still lacking for 
the Jain states.

\section{Tunneling across a quantum Hall strip}
\label{strip}

In this section, we return to the one-particle kernel and show how it can be
used in a familiar setting, namely, to study tunneling between the edges of a
quantum Hall strip. This will provide us with a microscopic understanding of
some of the statements which have been made in the literature, in particular,
the possibility of interference between tunneling due to multiple impurities.
\cite{chamon1} We believe that the approach taken here is a first step towards
connecting the bulk physics of the LLL to a boundary. A full understanding of
bulk-mediated fractional statistics properties will require an analysis
similar to the one presented here for {\it two-particle} properties and is
beyond the scope of this paper.

We begin by examining a system of non-interacting electrons with filling
fraction $\nu = 1$. We consider a
Hall strip which is infinitely extended along the $\hat x$-direction, and is
confined in the $\hat y$-direction by a potential $eV(y)$, as shown in Fig. 3.
For concreteness, we may take $V(y)$ to be simple harmonic. \cite{horsdal} We
choose the Landau gauge for the vector potential, so that the Hamiltonian is
\beq H ~=~ \frac{1}{2\mu} ~(p_x + \frac{eB}{c} y)^2 ~+~ \frac{p_y^2}{2\mu} ~+~
eV(y). \label{ham4} \eeq
The eigenstates of this Hamiltonian take the form
\beq \psi_{k,n} (x,y) ~=~ e^{ikx} ~f_{k,n} (y), \eeq
where $f_{k,n}$ satisfies the eigenvalue equation
\beq [\frac{p_y^2}{2\mu} ~+~ \frac{1}{2\mu} ~(\hbar k + \frac{eB}{c} y)^2 ~+~
eV(y)]~ f_{k,n} ~=~ E_{k,n} ~f_{k,n}. \eeq
The wave number $k$ is quantized in units of $2\pi /L_x$ if the system has
a length $L_x$; $k$ becomes a continuous variable as $L_x \to \infty$. If we
work at zero temperature and the Fermi energy of the electrons is $E_F$, only
those states will be occupied for which $E_{k,n} < E_F$. Let us assume that
the confining potential is sufficiently weak (i.e., it is much less than the
Landau level spacing $\hbar \omc$ in the region of interest), so that only
states with $n=0$ can be filled. As a function of $k$, the filled states lie
within some interval $k_1 < k < k_2$. Assuming that $V(y)$ is an even function
of $y$, we have $k_2 = k_F$ and $k_1 = - k_F$, and the corresponding wave
functions are centered about $y=-y_F$ and $y_F$ respectively, where
\beq y_F ~=~ k_F l^2, \label{yfkf} \eeq
$l$ being the Landau length.
(It is a well-known feature of states in the Landau gauge that the momentum
in the $\hat x$-direction and the position in the $\hat y$-direction are
correlated with each other). In the presence of the confining potential, the
energy levels in the LLL are no longer degenerate. If the confining potential
$eV(y)$ is weak, there is an effective electric field at the edge given by
${\cal E} = -(dV/dy)_{y=y_F}$. Then the electrons have a drift velocity given
by $v_F = c |{\cal E}/B|$. If $|{\cal E}/B| \ll 1$, we can show that $v_F =
(1/\hbar) (\partial E_{k,0} /\partial k)_{k=k_F}$. \cite{girvin} The
electrons have a negative velocity $-v_F$ at the upper edge at $y=y_F$,
and a positive velocity $v_F$ at the lower edge at $y=-y_F$.

\begin{figure}[t]
\ig[width=3.25in]{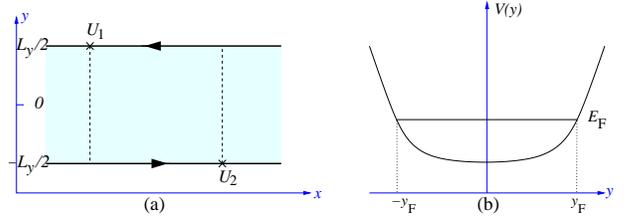}
\caption[]{(Color online) Quantum Hall state in a strip geometry of length
$L_x$ and width $L_y$ in the presence of a confining potential $V(y)$. The
states are filled up to a Fermi energy $E_F$, corresponding to a strip width
of $2y_F = L_y$. Tunneling across the strip can take place via impurities
denoted by $U_i$.}
\vspace{-0.1cm} \label{fig:Hallbar}
\end{figure}

In order to have tunneling between the two edges, it is necessary to break
the translation invariance in the $\hat x$-direction. As a simple model for
this, let us introduce an impurity potential of the form $U_1 (x,y) = U_1
\de (x-x_1) \de (y-y_1)$; we have introduced the subscript 1 because we
will later consider the effect of more than one impurity. Assuming that $U_1$
is small, we will use the Born approximation to compute the tunneling
amplitude produced by this impurity. Although the impurity potential has
non-zero matrix elements between states with all possible energies, we will
see that in the asymptotic regions $x \to \pm \infty$, the impurity causes
scattering only between pairs of states with the same energy. We consider
an electron coming in from the left in the state $\psi_{k_F,0}$ with energy
$E_F$. Under the lowest order Born approximation, we have
\bea \psi ({\vec r}) &=& \psi_{k_F,0} ({\vec r}) \non \\
& & - ~\int d^2 {\vec r}~' ~G ({\vec r}, {\vec r}~', E_F) ~U_1 ({\vec r}~')~
\psi_{k_F,0} ({\vec r}~'), \non \\
& & \label{born} \eea
where the retarded Green's function is given by
\beq G ({\vec r}, {\vec r}~', E) ~=~ \sum_n ~\int ~\frac{dk}{2\pi} ~ \frac{
\psi_{k,n} ({\vec r}) ~\psi_{k,n}^* ({\vec r}~')}{E_{k,n} ~-~ E ~-~ i \ep}.
\eeq
We recall that the Fourier transform of the above Green's function, $G
({\vec r}, {\vec r}~', t)$, vanishes for $t < 0$ and is equal to $i/\hbar$
times the one-particle kernel for $t> 0$.

We now evaluate Eq. (\ref{born}) using some approximations. First, we will
only consider contributions to the Green's function coming from states with
$n=0$ whose energies lie in the vicinity of $E_F$. For these states, we make
the linearized approximation
\bea E_{k,0} &=& E_F ~+~ \hbar v_F (k-k_F) ~~{\rm near} ~~k=k_F, \non \\
&=& E_F ~-~ \hbar v_F (k+k_F) ~~{\rm near} ~~k=-k_F, \label{ek} \eea
and we then extend the range of the integration variable $k$ up to
$\pm \infty$. Secondly, since the confining potential $eV(y)$ is weak, we
approximate the wave functions by
\bea \psi_{k,0} ({\vec r}) ~=~ e^{ikx} \left( \frac{1}{\pi l^2} \right)^{1/4}
\exp [- ~(y \pm y_F)^2 /(2l^2) ] \non \\
& & \label{psik} \eea
for $k$ close to $\pm k_F$ respectively. Eq. (\ref{born}) then gives
\bea \psi ({\vec r}) &=& \psi_{k_F,0} ~+~ r_L ~\psi_{-k_F,0} ~~{\rm for} ~~
x \to - ~\infty, \non \\
&=& t_L ~\psi_{k_F,0} ~~{\rm for} ~~ x \to \infty, \eea
where the reflection and transmission amplitudes are given by
\bea r_L &=& - ~\frac{iU_1}{\hbar v_F} ~e^{i2k_F x_1} \non \\
& & ~\times ~\left( \frac{1}{\pi l^2} \right)^{1/2} \exp [- ~(y_1^2 + y_F^2)/
l^2 ], \non \\
t_L &=& 1 ~-~ \frac{iU_1}{\hbar v_F} \left( \frac{1}{\pi l^2} \right)^{1/2}
\exp [- ~(y_1 + y_F)^2 /l^2] \non \\
& & \label{rl} \eea
to first order in $U_1$. The Gaussian factor appearing in $r_L$ is dependent on
$(y_1 + y_F)^2 + (y_1 - y_F)^2$, i.e., the sum of the squares of the distances
of the impurity from the two edges. The probability of tunneling between the
two edges is given by $|r_L|^2$. Similarly, for an electron coming in from the
right in the state $\psi_{-k_F,0}$ with energy $E_F$, Eq. (\ref{born}) gives
\bea \psi ({\vec r}) &=& \psi_{-k_F,0} ~+~ r_R ~\psi_{k_F,0} ~~{\rm for} ~~
x \to \infty, \non \\
&=& t_R ~\psi_{-k_F,0} ~~{\rm for} ~~ x \to - ~\infty, \eea
where
\bea r_R &=& - ~\frac{iU_1}{\hbar v_F} ~e^{-i2k_F x_1} \non \\
& & ~\times ~\left( \frac{1}{\pi l^2} \right)^{1/2} \exp [- ~(y_1^2 + y_F^2)/
l^2 ], \non \\
t_R &=& 1 ~-~ \frac{iU_1}{\hbar v_F} \left( \frac{1}{\pi l^2} \right)^{1/2}
\exp [- ~(y_1 - y_F)^2 /l^2]. \non \\
& & \eea
We can introduce a $2 \times 2$ scattering matrix which relates the
outgoing states to the incoming states,
\bea S &=& \left( \begin{array}{cc} r_L & t_R \\
t_L & r_R \end{array} \right). \eea
We find, as required, that $S$ is unitary to first order in $U_1$.

We can now consider the effects of several impurities which produce a
potential of the form
\beq U (x,y) ~=~ \sum_n ~U_n ~\de (x-x_n) ~\de (y-y_n). \eeq
To first order in the $U_n$, we get
\bea r_L &=& - ~\frac{i}{\hbar v_F} ~\sum_n ~U_n ~e^{i2k_F x_n} ~\exp
[- ~y_n^2 /l^2] \non \\
& & ~\times ~\left( \frac{1}{\pi l^2} \right)^{1/2} \exp [- ~ y_F^2 / l^2],
\label{rtot} \eea
and $r_R = - ~r_L^*$. We see that there are certain configurations
of two or more impurities which give zero reflection. We can understand
this using an Aharonov-Bohm phase argument in a simple case involving two
weak barriers, \cite{chamon1} as shown in Fig. 3. Consider two impurities
with equal strengths $U_1 = U_2$, and $y_1 = \pm y_2$. According to Eq.
(\ref{rtot}), $r_L$ vanishes if
\beq 2k_F (x_2 - x_1) ~=~ \frac{2y_F (x_2 - x_1)}{l^2} ~=~ (2n+1)~ \pi.
\label{ab} \eeq
Now, $2y_F (x_2 - x_1)$ is the area of the rectangle bounded by the two edges
and by the lines $x=x_1$ and $x=x_2$. If we imagine that the tunneling between
the edges occurs along the straight lines $x=x_1$ and $x=x_2$, Eq. (\ref{ab})
represents the Aharonov-Bohm phase enclosed by the edges and the tunneling
paths. The reflection amplitude vanishes if this phase is an odd multiple of
$\pi$.

Although the results obtained so far are for the case of electrons with
$\nu = 1$, we expect that certain aspects of the results will continue to
remain valid for quasi-particles with charge $\pm e/m$. The effective width
of the Hall strip, $2y_F$, is expected to be the same for electrons and for
quasi-particles. The forms of the wave function in Eq. (\ref{psik}) and
therefore of the tunneling amplitude in Eq. (\ref{rtot}) are also expected to
remain the same for quasi-particles, except that the factors of $y^2/l^2$ in
the exponentials will change to $y^2 /(ml^2)$ due to the charge of the
quasi-particles. Finally, the condition for the vanishing of the reflection
amplitude due to two impurities given in Eq. (\ref{ab}) will change to $2y_F
(x_2 - x_1) /(ml^2) = (2n+1) \pi$; this is again because the quasi-particle
charge is different from that of electrons, but the area of the rectangle
continues to be $2y_F (x_2 - x_1)$.

Since the exponentials appearing in the tunneling amplitudes $r_L$ and $r_R$
depend on the charge of the tunneling particle, the tunneling amplitude will
be much larger for quasi-particles with charge $\pm e/m$ than for electrons
with charge $e$. \cite{auerbach} The situation is of course complicated by the
fact that the effective tunneling amplitude is governed by a renormalization
group equation and therefore depends on the energy scale of interest.
\cite{wen1,wen2,wen3,kane2} What we have calculated in this section is the 
bare tunneling amplitude.

We have considered above the tunneling of a single particle, called $A$, from
one edge to another without taking into account its interaction with any other
particles. One can consider a case in which a second particle, called $B$, is
pinned to some fixed point inside the strip; let us denote the location of
this particle by $(x_B, y_B)$. If the two particles are, say, quasi-holes, the
expression given in Eq. (\ref{rtot}) for the tunneling due to several
impurities will need to be modified in order to take their fractional
statistics into account. For the reflection amplitude $r_L$ in (\ref{rtot}),
the contribution from an impurity which lies to the left of particle $B$
(i.e., $x_n < x_B$) would remain unchanged, while the contribution from an
impurity which lies to the right of $B$ ($i.e., x_n > x_B$) would carry an
extra phase of $e^{-i2\pi \al}$. This is because the tunneling path
corresponding to the second impurity has an extra phase compared to the
tunneling path corresponding to the first impurity; this is given by the
phase of one anyon going around another in the clockwise direction.

The arguments given above explain why the total tunneling amplitude for one
particle will oscillate if one varies either the magnetic field or the number
of other particles lying within the rectangle formed by two impurities. This
provides a microscopic justification for some of the statements made in Ref.
\onlinecite{chamon1}.

The calculations in this section can be generalized to the case of quantum
Hall systems in which more than two edges come close to each other in certain
regions \cite{safi,vishveshwara1,kim,grosfeld,oshikawa,das,comforti}. We
expect that the statement made after Eq. (\ref{rl}), namely, that an
impurity-induced tunneling amplitude between two edges depends on the
sum of the squares of the distances of the impurity from the two edges,
will remain valid in general.

\section{Correlations along edges of a quantum Hall system}

In this section, we discuss the two-point correlators along the edges of a
quantum Hall system. We will consider two possible geometries, namely, a long
strip and a circular droplet. Since our main aim is to show the effects of
different geometries, we restrict our attention to the case of non-interacting
electrons with $\nu = 1$ for simplicity.

Let us first consider the strip geometry. We assume a confining potential
as in Sec. \ref{strip}; hence the states in the LLL are labeled only by the
wave number $k$ in the $\hat x$-direction. The ground state $|G \rangle$ of
the electrons is taken to be one in which all states are filled up to a
Fermi energy $E_F$ which corresponds to $k=\pm k_F$. For simplicity, we
ignore the effect of the confinement when writing down the wave functions,
\bea \psi_{k,0} ({\vec r}) ~=~ e^{ikx} \left( \frac{1}{\pi l^2} \right)^{1/4}
\exp [- ~(y + k l^2)^2 /(2l^2)]. \non \\
& & \eea
We will compute the correlator
\beq C({\vec r}_1; {\vec r}_2; t) ~=~ \langle G | \Psi^\dg ({\vec r}_1, 0)
\Psi ({\vec r}_2, t)| G \rangle, \label{corr1} \eeq
where the second quantized field $\Psi$ is given by
\beq \Psi ({\vec r}, t) ~=~ \sum_n ~\int ~\frac{dk}{2\pi} ~c_{k,n} ~\psi_{k,n}
({\vec r}) e^{-iE_{k,n} t/\hbar} , \eeq
and $~\{ c_{k,n}, c_{k',n'}^\dg \} = 2\pi \de (k-k') \de_{n,n'}$.
For ${\vec r}_1 = (0,y_1)$ and ${\vec r}_2 = (x,y_2)$, we obtain the
expression for the equal-time correlator
\bea C(0,y_1; x,y_2; 0) &=& \int_{-k_F}^{k_F} ~\frac{dk}{2\pi} ~
e^{ikx} ~\left( \frac{1}{\pi l^2} \right)^{1/2} \non \\
& & ~~~\times ~\exp [- ~(y_1 + k l^2)^2 /(2l^2)] \non \\
& & ~~~\times ~\exp [- ~(y_2 + k l^2)^2 /(2l^2)]. \non \\
& & \eea
It is convenient to define the dimensionless variables $X = x/l$ and $Y_F =
k_F l$. We now consider the following cases.

\noi (i) The two points are on the same edge, say the lower edge $y_1 = y_2 = -
y_F$, where $y_F$ is related to $k_F$ through Eq. (\ref{yfkf}). After shifting
and rescaling $k$ to be dimensionless, we get \beq C(0,-y_F; x,-y_F; 0) ~=~
\frac{1}{2\pi^{3/2} l^2} ~\int_{-2Y_F}^0 dk ~ e^{- k^2 + ikX}. \label{corr2}
\eeq If the strip is much wider than the Landau length, namely, $Y_F \gg 1$,
the integral in Eq. (\ref{corr2}) is equal to $(1/2) e^{-X^2 /4} [\sqrt{\pi} -
iX {}_1F_1 (1/2; 3/2; X^2/4)]$. \cite{gradshteyn} For $X^2/4 \to \infty$, this
goes to zero as $-i /X$. \cite{abramowitz} The last result can be obtained in a
different way. In the limit of large $|X|$, the integral only gets a
contribution from small values of $k$. The asymptotic value of the integral
therefore does not depend on how the integrand is cut off at large values of
$|k|$. If we choose a cut-off such as $e^{\ep k}$, where we eventually take
$\ep \to 0^+$, the integral can be done analytically and we get $-i/X$.

\noi (ii) The two points are on opposite edges, $y_1 = - y_F$ and $y_2 = y_F$.
After rescaling, we get
\beq C(0,-y_F; x,y_F; 0) ~=~ \frac{1}{2\pi^{3/2} l^2} ~e^{-Y_F^2}~
\int_{-Y_F}^{Y_F}~ dk ~e^{- k^2 + ikX}. \eeq
If the strip is very wide, the limits of the integration can be replaced by
$\pm \infty$, and we find that the correlator goes as $e^{-X^2/4}$. On the
other hand, if the strip width is of the order of the Landau length, i.e.,
$Y_F \sim 1$, the magnitude of the correlator exhibits oscillations and, in
fact, vanishes for a discrete set of values of $X$. For small values of $Y_F$,
the wavelength of the oscillations is given by $\pi /Y_F$. Oscillations also
occur, although they are less prominent, if the two points lie on the same
edge and $Y_F \sim 1$. Fig. 4 shows these oscillations in the correlator
(given in units of $1/(2\pi^{3/2} l^2)$) for $Y_F = 0.5$.

The results discussed above show that the usual discussions of edge states and
bulk physics have to be modified in the case of highly confined geometries.
This needs to be kept in mind when one is trying to understand the results of
experiments performed under such conditions.

\begin{figure}[t]
\ig[width=2.5in]{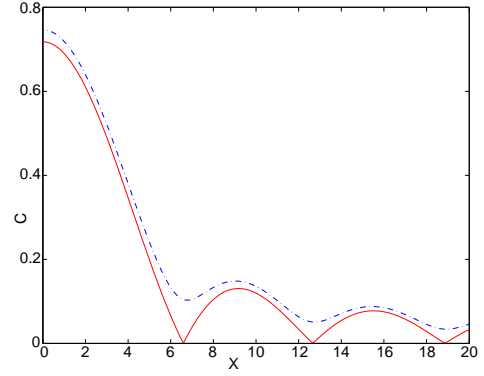}
\caption[]{(Color online) Magnitude of the two-point correlator versus the
separation $X$, for $Y_F = 0.5$. The solid and dash-dot lines show the cases
where the two points lie on opposite edges and on the same edge respectively.}
\vspace{-0.3cm} \label{fig:correlator}
\end{figure}

Let us now consider a circular geometry with a confining potential $V(r)$
which we will not explicitly specify. We choose the symmetric gauge, and use
the LLL wave functions given in Eq. (\ref{wave1}). In the presence of the
confining potential, we take the ground state to be one in which all states
from $n =0$ to $N$ are filled, where we assume that $N \gg 1$. The radius of
the system is given by $r_F = l \sqrt{2 N}$. Let us take the two points
${\vec r}_1,~ {\vec r}_2$ to lie on the circumference, with ${\vec r}_1 = r_F
(1,0)$ and ${\vec r}_2 = r_F (\cos \phi, \sin \phi)$. We define a dimensionless
radius $R = r_F /l = \sqrt{2N}$. The equal-time correlator is then given by
\beq C({\vec r}_1; {\vec r}_2; 0) ~=~ \frac{1}{2\pi l^2} ~\sum_{n=0}^N ~
\frac{1}{n!} ~\left(\frac{R^2 e^{i\phi}}{2}\right)^n ~\exp [-\frac{1}{2} R^2].
\label{cn} \eeq
For large values of $N$, we can evaluate the sum using saddle point methods.
\cite{bender} We first expand the terms in Eq. (\ref{cn}) around $n=N$ up to
second order in $n-N$; this yields
\beq C({\vec r}_1; {\vec r}_2; 0) ~\simeq~ \frac{e^{iN \phi}}{2\pi^{3/2} l^2
\sqrt{2N}}~ \sum_{n=0}^N e^{i(n-N)\phi - (n-N)^2/2N}. \label{circle} \eeq
If $\phi \ll \pi$, we can replace the sum in Eq. (\ref{circle}) by an integral.
Introducing the quantity $k=(n-N)/\sqrt{2N}$ which becomes a continuous
variable as $N \to \infty$, we find that
\beq C({\vec r}_1; {\vec r}_2 ;0) ~\simeq~ \frac{e^{iN \phi}}{2\pi^{3/2}
l^2}~ \int_{-\infty}^0~ dk ~e^{- k^2 + ik R \phi}. \label{corr3} \eeq
This has exactly the same form as in Eq. (\ref{corr2}), up to a phase
$e^{iN\phi}$, with $X$ being equal
to the arc length $R \phi$. On the other hand, if $\phi \sim \pi$, we cannot
replace the sum in Eq. (\ref{circle}) by an integral. However, the sum is
then dominated by small values of $n-N$, and we can replace the Gaussian
cut-off $e^{- (n-N)^2/2N}$ by $e^{\ep (n-N)}$, where we eventually take
$\ep \to 0^+$. On summing up the series, we obtain \cite{jain1,wen3}
\beq C({\vec r}_1; {\vec r}_2 ;0) ~\simeq~ - ~\frac{e^{iN \phi}}{2\pi^{3/2}
l^2} ~\frac{ie^{i\phi /2}}{2R \sin (\phi /2)}. \label{corr4} \eeq
For $\phi \to 0$ but $R \phi \gg 1$, Eq. (\ref{corr4}) agrees with the
result one obtains from Eq. (\ref{corr3}).

Finally, let us discuss the finite time behavior of the correlators for the
cases when the two points lie on the same edge of a strip, or on the
circumference of a circle. At large separations, i.e., $x \to \infty$ for the
strip or $r_F \phi \to \infty$ for the circle, the contribution to the
correlator mainly comes from states close to the Fermi energy. For such
states, we can use the linearized approximation for the dispersion, $E_{k,0}
= \hbar v_F (k-k_F)$ near the lower edge of the strip, or $E_n = \hbar v_F
(n-N)/r_F$ near the circumference of the circle; while writing these down, we
have re-defined the Fermi energy so that it is zero at the edge. We then see
that the finite-time correlator can be obtained from the equal-time correlator
by replacing $x$ by $x-v_F t$ at the lower edge of the strip, or $r_F \phi$
by $r_F \phi - v_F t$ at the circumference of the circle.

The calculations in this section are valid when the filling fraction $\nu$
is equal to 1. For $\nu = 1/m$, it is known that the two-point correlators
for electrons and quasi-particles decay as non-trivial powers of the
separation which can be found using the technique of bosonization
\cite{wen3,stone,gogolin,rao,giamarchi}. However, bosonization only works
at distances which are large compared to the Landau length. Hence, one may
require other techniques to study short-distance properties for $\nu = 1/m$.

\section{Summary}

In this paper, we have studied the one- and two-particle kernels of charged
particles moving in a strong magnetic field. We have shown that the
two-particle kernel in the bulk of a quantum Hall system contains information
about two important properties of the particles. Namely, the kernel vanishes
for a discrete set of geometries of the initial and final positions in a way
which is determined by the exchange statistics; the kernel also vanishes as
the particles approach each other with a power law which is related to a
generalized exclusion statistics. These angular and distance dependences of
the kernel should be observable in the correlations of two-particle tunnelings
in appropriate gate-defined quantum Hall geometries. We have proposed an
experiment which can use the vanishing of the kernel to determine the charge
and fractional statistics of the particles. Our analysis is expected to work 
best for fractional quantum Hall states lying in the Laughlin sequence with 
the filling fraction being given by the inverse of an odd integer. For such
states, a lowest Landau level treatment is adequate, and the charge and 
statistics parameter of quasi-particles are equal to each other.

The one-particle kernel can be used to study impurity-induced tunneling
between different edges of a quantum Hall strip which contains some impurities.
Here too we find that certain arrangements of the impurities give rise to a
vanishing tunneling amplitude. Finally, we have studied the two-point
correlator along different edges of a quantum Hall strip and droplet. We find
that the correlator has a rich structure if the separation between the two
points and the width of the strip are of the order of the Landau length.

To conclude, we see that geometry plays an important, and sometimes
surprising, role in the propagation of particles through the bulk of a
quantum Hall system. A complete understanding of experiments which measure
one- and two-particle properties therefore requires us to take into account
all the geometrical aspects of the problem.

\section*{Acknowledgments}

We would like to acknowledge Assa Auerbach, Gordon Baym, Eduardo Fradkin,
Moty Heiblum and Jainendra Jain for illuminating discussions. This work was
supported by the UIUC Department of Physics and by the NSF under Grants No. 
DMR 06-03528 and DMR 06-44022. Work in India was supported by DST under the 
project SR/S2/CMP-27/2006.

\appendix

\section{One-particle kernel in a magnetic field}

To derive the kernel given in Eq. (\ref{k1mag1}), we start with the action
\beq S ~=~ \int_0^t ~dt' ~[~ \frac{\mu}{2} ({\dot x}^2 ~+~ {\dot y}^2) ~+~
\frac{eB}{2c} (x {\dot y} ~-~ y {\dot x}) ~]. \label{act} \eeq
We find the Euler-Lagrange equations of motion arising from this action, and
solve them using the boundary conditions ${\vec r} (t'=0) = {\vec r}_i$ and
${\vec r} (t'=t) = {\vec r}_f$. Substituting the solution in Eq. (\ref{act}),
we obtain the classical action
\bea S_{cl} ({\vec r}_f; {\vec r}_i; t) &=& \frac{\mu\omc \cot (\omc t/2)}{4}~
({\vec r}_f - {\vec r}_i)^2 \non \\
& & +~ \frac{\mu\omc}{2} ~{\hat z} \cdot {\vec r}_i \times {\vec r}_f.
\label{classact} \eea
Since the action is quadratic in $x$ and $y$, a path integral argument shows
that the kernel must be of the form \cite{feynman}
\beq K_1 ({\vec r}_f; {\vec r}_i; t) ~=~ f(t) ~e^{iS_{cl} ({\vec r}_f;
{\vec r}_i; t)/\hbar}, \label{ft} \eeq
where $f(t)$ is independent of ${\vec r}_i$ and ${\vec r}_f$. Next, we know
that the kernel is given by the sum
\beq K_1 ({\vec r}_f; {\vec r}_i; t) ~=~ \sum_n ~\psi_n ({\vec r}_f) ~\psi_n^*
({\vec r}_i) ~e^{-iE_n t/\hbar}, \eeq
and therefore satisfies the equation
\beq [~i \hbar ~\frac{\partial}{\partial t} ~-~ H_f ~] ~K_1
({\vec r}_f; {\vec r}_i; t) ~=~ 0, \eeq
where $H_f$ is the Hamiltonian in Eq. (\ref{ham1}) written in terms of
${\vec r}_f$ and $\partial /\partial {\vec r}_f$. This leads to the following
first order differential equation for the function $f(t)$ in Eq. (\ref{ft}),
\bea \frac{df}{dt} &=& - ~\frac{i}{\hbar} ~\left[ \frac{\partial S_{cl}}{
\partial t} ~+~ e^{-iS_{cl}/\hbar} ~H_f ~e^{iS_{cl}/\hbar} \right] ~f \non \\
&=& -~ \frac{\omc}{2} ~\cot (\frac{\omc t}{2}) ~f. \eea
The solution of this equation is
\beq f(t) ~=~ \frac{A}{\sin (\omc t/2)}. \eeq
Finally, the constant $A$ can be fixed by demanding that the kernel should
reduce to the free particle result in Eq. (\ref{k1}) as $\omc \to 0$. This
yields $A= \mu\omc /(i 4\pi \hbar)$. Putting all this together gives Eq.
(\ref{k1mag1}).

\section{Alternative formalisms for a two-anyon system}

Consider a system of two anyons with a relative coordinate ${\vec r}$ and mass
$\mu /2$; let us assume that there are no external fields or potentials. One
can proceed in two different ways using the polar coordinates $(r,\phi)$.
\cite{wilczek2,forte,lerda,khare,myrheim,ouvry}

In the first formalism, the Lagrangian and Hamiltonian are given by
\bea L &=& \frac{\mu}{4} ~({\dot r}^2 ~+~ r^2 {\dot \phi}^2), \non \\
{\rm and} ~~H &=& \frac{1}{\mu} ~(p_r^2 ~+~ \frac{p_\phi^2}{r^2}),
\label{form1} \eea
where $p_\phi = - i \hbar \partial /\partial \phi$. In this case, the wave
functions are taken to be multi-valued with the property that $\psi (\phi +
\pi) = e^{i\pi \al} \psi (\phi)$; this is satisfied if the angular dependence
of the wave function is of the form $e^{i(2p + \al) \phi}$, where $p$ is an
integer.

In the second formalism, the Lagrangian and Hamiltonian are given by
\bea L &=& \frac{\mu}{4} ~({\dot r}^2 ~+~ r^2 {\dot \phi}^2) ~-~ \al
{\dot \phi}, \non \\
{\rm and} ~~H &=& \frac{1}{\mu} ~\left( p_r^2 ~+~ \frac{(p_\phi + \al)^2}{r^2}
\right). \label{form2} \eea
In this case, the wave functions are single-valued; the angular dependence
is of the form $e^{i2p\phi}$, where $p$ is an integer. The energy levels are
of course the same in the two formalisms.

Let us now consider studying the problem using path integrals. If we exchange
the two particles in an anticlockwise sense, the coordinate $\phi$ changes by
$\pi$. In the second formalism, the path integral picks up a phase of
$e^{-i \pi \al}$ due to the last term in the Lagrangian in Eq. (\ref{form2}).
For a clockwise exchange of the two particles, $\phi$ changes by $-\pi$ and
the path integral picks up a phase of $e^{i \pi \al}$. We thus see that the
exchange phases picked up by the wave function in the first formalism and by
the path integral in the second formalism have opposite signs.

\section{Charge and statistics of quasi-holes in a fractional quantum Hall
system}

The charge and statistics of quasi-holes in a quantum Hall system with filling
fraction equal to $1/m$, where $m$ is an odd integer, has been derived in
Refs. \onlinecite{laughlin2,halperin,arovas}. Briefly, the derivation is based
on the Laughlin variational wave functions for the ground state $\psi_0$ with
no quasi-holes, $\psi_1 (\eta_1)$ describing one quasi-hole located at
$\eta_1$ (using complex notation), and $\psi_1 (\eta_1, \eta_2)$ describing
two quasi-holes located at $\eta_1$ and $\eta_2$. These are given by
\bea \psi_0 &=& \prod_{i<j} ~(z_i ~-~ z_j)^m ~\exp [-\frac{1}{4} ~\sum_i ~
|z_i|^2], \non \\
\psi_1 (\eta_1) &=& \prod_i ~(z_i ~-~ \eta_1) ~\psi_0 , \non \\
\psi_2 (\eta_1, \eta_2) &=& \prod_i ~(z_i ~-~ \eta_1) (z_i ~-~ \eta_2) ~
\psi_0 , \eea
where $z_i$ denote the locations of the electrons. We compute the phase picked
up by the wave function $\psi_1 (\eta_1)$ when $\eta_1$ is taken around a large
anticlockwise loop enclosing an area $A$. Equating that to the Aharonov-Bohm
phase of a particle of charge $e^*$ moving in a magnetic field, one finds that
the quasi-hole has charge $-e/m$. Then we consider the phase picked up by
$\psi_2 (\eta_1, \eta_2)$ when $\eta_1$ is taken around a large anticlockwise
loop of area $A$ which encloses the quasi-hole at $\eta_2$. This phase is given
by a sum of the Aharonov-Bohm phase due to the magnetic field and twice the
exchange phase; the latter arises because taking one quasi-hole around another
gives a phase which is twice the phase picked up when the two quasi-holes are
exchanged. We thus discover that the exchange phase is $\pi /m$. Now, the
wave function $\psi_2$ is clearly single-valued. Following the arguments given
in Appendix B, we therefore identify $\al = - 1/m$.

Next, let us consider the wave function of two quasi-holes in the LLL, where
the quasi-holes are now to be thought of as two `elementary' objects moving
in a vacuum, not as collective excitations of many electrons as described by
$\psi_2$ above. Since the sign of the quasi-hole is opposite to the sign of an
electron, the wave function must be a function of $z^*$ times a Gaussian. The
fact that $\al = -1/m$ now implies that if we use a multi-valued wave function,
the dependence of the wave function on the relative coordinate must be of the
form $(z_1^* - z_2^*)^{2p+1/m}$, where $p$ is a non-negative integer.

Although the wave function of quasi-holes is a function of $z^*$ in the LLL
(if the wave function of electrons is a function of $z$), we have taken the
quasi-hole wave functions to be functions of $z$ and changed $\al \to 1/m$
in the main body of the paper in order to use the same notation everywhere.

\section{Quasi-hole exclusion statistics}

We present here a state counting argument to show that quasi-holes in the LLL
exhibit exclusion statistics \cite{haldane,veigy,johnson,su}. The polynomial
part of a state of $N_q$ quasi-holes in a sea of $N_e$ electrons at filling
fraction $1/m$ is given by
\bea & & \psi(z_1,\ldots, z_{N_e};\zeta_1,\ldots, \zeta_{Nq}) \non \\
& & ~=~ \prod_{i=1}^{N_e} ~\prod_{a=1}^{N_q} ~(z_i-\zeta_a) ~\prod_{i<j} ~
(z_i-z_j)^m. \eea
This is a polynomial, in any individual $z_i$, of maximum degree
\beq N_\phi ~=~ m(N_e-1) + N_q ~=~ \frac{eB}{2\pi \hbar c} ~(\hbox{Area}).\eeq
Note that we have to keep this area (and not $N_e$) fixed as we count the
dimension of the quasi-hole Hilbert space. Now
\bea \prod_{i=1}^{N_e} ~(z_i-\zeta) &=& e_{N_e} ~-~ e_{N_e-1} \zeta ~+~
e_{N_e-2} \zeta^2 ~+~ \cdots \non \\
& & +~ (-1)^{N_e} \zeta^{N_e}, \eea
where $e_n(z_i)$ denote the elementary symmetric functions in the $N_e$
variables $z_i$. They can be represented as single-column Young diagrams with
at most $N_e$ boxes in the column.

A general linear combination of quasi-hole states is therefore a Laughlin
factor multiplied by a linear combination of products of $N_q$ such
elementary symmetric functions. The total number of such functions
\footnote{Note that the products $e_{\lambda_1}e_{\lambda_2}\cdots$, with
$N_e\ge\lambda_1\ge \lambda_2$, etc, are linearly independent, and they span
the space of symmetric polynomials of degree $|\lambda|\equiv \sum \lambda_i$;
this is a standard result in the theory of symmetric functions.}
is given by the number of partitions that can be fitted into an $N_e$-by-$N_q$
rectangle. This number is the number of positive-going random walks on an
integer lattice from $(0,0)$ to $(N_e,N_q)$. It is therefore the coefficient
of $x^{N_e}y^{N_m}$ in the expansion of $(x+y)^{N_e+N_m},$ which is given by
\beq {{N_e+N_q}\choose {N_q}} ~=~ {{D+N_q-1}\choose {N_q}}. \eeq
In the second form, we have eliminated $N_e$ in favor of $N_\phi$ and defined
\beq D_q ~=~ \frac {N_\phi} m ~-~ \frac 1 m(N_q -1) ~+~ [2-1/m], \eeq
where $[r]$ denotes the integer part of $r$. Ignoring the term $[2-1/m]$
which is thermodynamically insignificant, \cite{su} we see that the expression
for $D$ is of the Haldane form \cite{haldane}
\beq D ~=~ \frac{N_\phi}m ~-~ g(N_q-1), \eeq
with the exclusion parameter $g=1/m$. ($g$ is 1 for fermions and 0 for bosons).
Note that $N_\phi/m$ is the number of single particle states available to a
charge $e/m$ particle in a region threaded by $N_\phi$ electron flux units.

\end{document}